\documentclass[preprint,12pt,authoryear]{elsarticle}
\usepackage[letterpaper,margin=1in]{geometry}
\usepackage{setspace}
\usepackage{endnotes}
\usepackage{bbm}
\usepackage[T1]{fontenc}
\usepackage{lmodern}
\usepackage{algorithm, calc, algpseudocode, mdframed, caption}
\usepackage{booktabs}
\usepackage{enumitem}
\usepackage{graphicx}
\usepackage{placeins}
\usepackage{hyperref}
\usepackage[scr=stixtwofancy]{mathalpha}
\usepackage{xcolor}
\usepackage{threeparttable}
\usepackage{amssymb}
\usepackage{amsmath}
\usepackage{comment}
\usepackage{etoolbox}
\let\originalbibitem\bibitem
\renewcommand{\bibitem}[2][]{%
	\originalbibitem[#1]{#2}%
	\ifstrequal{#2}{FrenchData2026}{[dataset]\space}{}%
}
\usepackage[normalem]{ulem}
\hypersetup{hidelinks}
\journal{Journal of Financial Econometrics}
\colorlet{shadecolor}{gray!25}

\newcommand{\stepoutput}[1]{%
	\Statex
	\fcolorbox{blue!30!black}{blue!3}{%
		\parbox{\dimexpr\linewidth-3.2em\relax}{%
			\textbf{\textcolor{blue!40!black}{Result:}} #1%
		}%
	}%
}
\newcommand{\newoperator}[3]{\newcommand*{#1}{\mathop{#2}#3}}
\newcommand{\renewoperator}[3]{\renewcommand*{#1}{\mathop{#2}#3}}
\renewoperator{\Re}{\mathrm{Re}}{\nolimits}
\renewoperator{\Im}{\mathrm{Im}}{\nolimits}
\makeatletter
\newcommand{\rd}{\@ifnextchar^{\DIfF}{\DIfF^{}}}
\def\DIfF^#1{%
   \mathop{\mathrm{\mathstrut d}}%
   \nolimits^{#1}\gobblespace}
\def\gobblespace{\futurelet\diffarg\opspace}
\def\opspace{%
   \let\DiffSpace\!%
   \ifx\diffarg(%
   \let\DiffSpace\relax
   \else
   \ifx\diffarg[%
   \let\DiffSpace\relax
   \else
   \ifx\diffarg\{%
   \let\DiffSpace\relax
   \fi\fi\fi\DiffSpace}

\newcommand{\diag}{\operatorname{diag}}

\newoperator{\ip}{\mathrm{int}}{\nolimits}


\newcommand{\beq}{\begin{equation}}
\newcommand{\eeq}{\end{equation}}
\newcommand{\bal}{\begin{align*}}
\newcommand{\eal}{\end{align*}}

\newcommand{\bvec}{\begin{pmatrix}}
\newcommand{\evec}{\end{pmatrix}}
\newcommand{\bmat}{\begin{bmatrix}}
\newcommand{\emat}{\end{bmatrix}}

\newcommand{\bsmat}{\begin{smallmatrix}}
\newcommand{\esmat}{\end{smallmatrix}}

\newcommand{\squad}{\hspace{\algorithmicindent}}
\newcommand{\Tau}{\mathcal{T}}

\let\footnote\endnote
\newcommand{\alttext}[1]{%
	\par\medskip
	\noindent
	\begin{minipage}{\linewidth}
		\footnotesize
		\raggedright
		\noindent Alt text: #1\par
	\end{minipage}
	\par
}

\begin{document}
	\doublespacing
	\pagenumbering{arabic}
	\setcounter{page}{1}
	
	\thispagestyle{plain}
	\begin{center}
		{\Large\bfseries Exploiting Common Volatility Dynamics in
			High-Dimensional Portfolio Selection\par}
		\vspace{1.5em}
		Dominik Ammon \qquad Julian Heinz\\
		\vspace{0.6em}
		University of Regensburg\\
		Universit\"atsstra\ss e 31, 93053 Regensburg, Germany
	\end{center}
	\vspace{1.5em}
	\noindent\textbf{Corresponding author:} Julian Heinz\\
	\noindent\textbf{Address:} University of Regensburg,
	Universit\"atsstra\ss e 31, 93053 Regensburg, Germany\\
	\noindent\textbf{Email:} Julian.Heinz@ur.de\\
	\noindent\textbf{Telephone:} +49 941 943 2721
	\par
	\noindent\textbf{Coauthor email:} Dominik.Ammon@ur.de
	\clearpage

\noindent\textbf{Abstract}\par\medskip 
We study how commonalities in monthly realized stock volatilities can be exploited
for high-dimensional covariance forecasting and portfolio selection. We compare
latent factor models with ARMA models that pool dynamic parameters within
pre-specified groups of stocks. We show that pooled ARMA models admit an exact
dynamic factor representation with group-specific dynamics and stock-specific
loadings. The implied static factor representation involves group-specific factors that
need not be pervasive. Combining volatility forecasts with realized dynamic
conditional correlations and nonlinear shrinkage, we evaluate out-of-sample
minimum variance portfolios of S\&P 500 constituents. Pooling-based approaches
consistently outperform the alternatives across portfolio constraints and model specifications.
\par\medskip
\noindent\textbf{Keywords:} Portfolio selection; High dimensions;
Volatility modeling and forecasting; Factor models; Time-series models
\par\noindent\textbf{JEL classification:} C53; C58; G11
\bigskip

\label{sec:introduction}
Managing risk in portfolio allocation becomes particularly challenging as the 
investment universe expands. The number of volatility and dependence processes 
to be estimated grows rapidly with the cross-sectional dimension, 
whereas the available time-series histories remain comparatively short, 
rendering unrestricted models highly vulnerable to estimation errors.
The pronounced comovement of stock volatilities offers a way to reduce this burden. 
Individual volatilities respond jointly to market-wide and sector-specific shocks 
and often display similar persistence patterns \citep{Bollerslev_2018,Engle2023}. 
We therefore consider two ways of exploiting this commonality: latent volatility factors and the pooling of volatility dynamics across stocks.

The two approaches impose parsimony in different ways. Factor models summarize
cross-sectional comovement through few latent components while
allowing stocks to have heterogeneous exposures to those components.
Parameter-pooling models retain the individual volatility series but require stocks across the market or within pre-specified groups to share the same dynamic coefficients.
Factor models therefore restrict the cross-sectional representation of
volatility, whereas pooling restricts heterogeneity in the time-series
dynamics.

We examine these models using monthly realized volatilities for a large
cross-section of S\&P 500 constituents. Each rolling estimation window contains
several hundred stocks but only 120 monthly observations, making the balance
between flexibility and precision important. Central to this study are static factor models (SFMs) based on principal components \citep{Stock2002}, the generalized
dynamic factor model (GDFM) of \citet{Forni2015,Forni2017}, and pooled ARMA models. Our pooling specifications build on the panel approach of \citet{Bollerslev_2018} and involve jointly estimating a set of dynamic parameters from multiple time series.
We consider market-wide pooling, which restricts dynamic parameters to be identical across all stocks, and industry-level pooling, which allows them to differ across industries while remaining identical for stocks within the same industry.
We focus on ARMA models because they offer a parsimonious way to approximate the long-memory dynamics that characterize realized volatility \citep{Andersen2003,Corsi2009,Hartl2022}.

Our first contribution is to clarify the structural relationship between these pooled models and factor models. 
We show that when innovations contain common shocks, the model admits an exact dynamic factor representation. Factor loadings remain stock-specific, while stocks within the same pooling group share a dynamic filter for both the common and idiosyncratic components. Relative to the GDFM, pooling gives rise to a bias-variance trade-off. Replacing stock-specific dynamic loading filters with a small number of group-specific filters reduces the parameter burden and estimation uncertainty at the cost of specification bias when volatility dynamics are heterogeneous.

The comparison with the SFM is less clear-cut. The pooled ARMA also admits a static factor representation, although under group-specific pooling its filtered factors need not be pervasive. Conversely, the empirical SFM retains a small number of pervasive principal-component factors while allowing stock-specific forecasting relationships and separate idiosyncratic dynamics. Thus, the pooled ARMA and SFM impose different restrictions and are not necessarily nested within one another.

Our second contribution is to evaluate the economic consequences of these
alternative restrictions in a high-dimensional covariance forecasting
framework. Each volatility forecast is combined with the same realized dynamic
conditional correlation model. The correlation specification builds on the dynamic conditional correlation (DCC) framework of \citet{Engle2002}, incorporates realized correlation as in \citet{Bollerslev_2020},
and regularizes its long-run target
using nonlinear shrinkage \citep{Engle2019,Ledoit2022}. Holding the correlation
model fixed allows the comparison to focus on how the competing volatility
models exploit cross-sectional information. We evaluate the resulting
covariance forecasts through out-of-sample global minimum variance (GMV) portfolios, allowing us to assess forecast performance directly in economic terms.

This paper's focus on commonality in volatility distinguishes it from much of the literature on high-dimensional portfolios. Many existing factor-based estimators exploit common variation in stock returns to regularize covariance matrix estimates \citep{Fan2013,DeNard2021,Daniele2025,Shi2025}. Instead, we shift the focus from common variation in returns to common variation in realized volatility of returns. Importantly, these forms of commonality need not coincide, as shown, for example, in \citet{Herskovic_2016}, \citet{Barigozzi2017, Barigozzi2020}, and \citet{Bekaert_2025}.
Our analysis is closely related to the literature on factor models for large
panels of volatility measures. \citet{Luciani2015} develop an approximate
dynamic factor model for realized volatilities that accommodates 
comovement and long-memory features. \citet{Barigozzi2017,Barigozzi2020}
develop a two-stage generalized dynamic factor approach to the joint analysis
of returns and volatilities. Related work considers long-range dependence in
factor dynamics \citep{Ergemen2023a} and multiplicative factor structures for
volatility \citep{Ding2025}. We complement this literature by comparing latent
volatility factors with parameter pooling and by characterizing the
restrictions that connect the two approaches.

The empirical results favor strong regularization of volatility dynamics. The market-pooled ARMA produces the lowest out-of-sample portfolio volatility in the baseline GMV and long-only settings, while the industry-pooled specification performs best under leverage- and turnover-constrained portfolios, marginally outperforming the market-pooled specification. In the baseline setting, the market-pooled ARMA reduces annualized portfolio standard deviation by \(20.2\%\) relative to the unconditional nonlinear-shrinkage benchmark. The results are robust to longer estimation windows, alternative correlation specifications, and higher-order ARMA dynamics, and they hold across subperiods.

 To better understand differences in portfolio performance, we analyze the underlying volatility forecasts, focusing on their cross-sectional correlation with realized volatilities and distinguishing errors in the overall volatility level from those that distort relative differences across stocks. The results are consistent with a bias-variance trade-off: the greater flexibility of factor models can reduce specification bias, while the parsimony of pooling can improve estimation precision.

The remainder of the paper is organized as follows. Section~\ref{Ch:Methodology} presents the volatility forecasting methodology and establishes the relationship between parameter pooling and factor representations. Section~\ref{Ch:CovarianceForecasting} describes the construction of conditional covariance matrices and introduces benchmark estimators. Section~\ref{Ch:empirical_results} describes the data and presents the empirical results. Section~\ref{Ch:Conclusion}
concludes.

\section{Realized Volatility and Forecasting}
\label{Ch:Methodology}
Throughout the paper, we follow the convention of representing one month by
21 consecutive trading days, indexed by $k=1,\ldots,21$. We use distinct time indices to distinguish between out-of-sample forecast months, indexed by $t=1,\ldots,T$, and in-sample months used for estimation, indexed by $\tau=1,\ldots,\Tau$.

\subsection{Commonalities in realized volatility}
\label{Ch:RVCommonality}

Let $r_{i,t,k}$ denote the daily log return of asset
$i=1,\ldots,N_t$, where $N_t$ denotes the investable universe in month $t$.
We define monthly realized volatility as
\begin{equation}
	RV_{i,t}
	=
	\left(
	\sum_{k=1}^{21}
	r_{i,t,k}^{2}
	\right)^{1/2}.
	\label{eq:RV}
\end{equation}

\begin{figure}[h!]
	\centering
	\includegraphics[width=\linewidth]{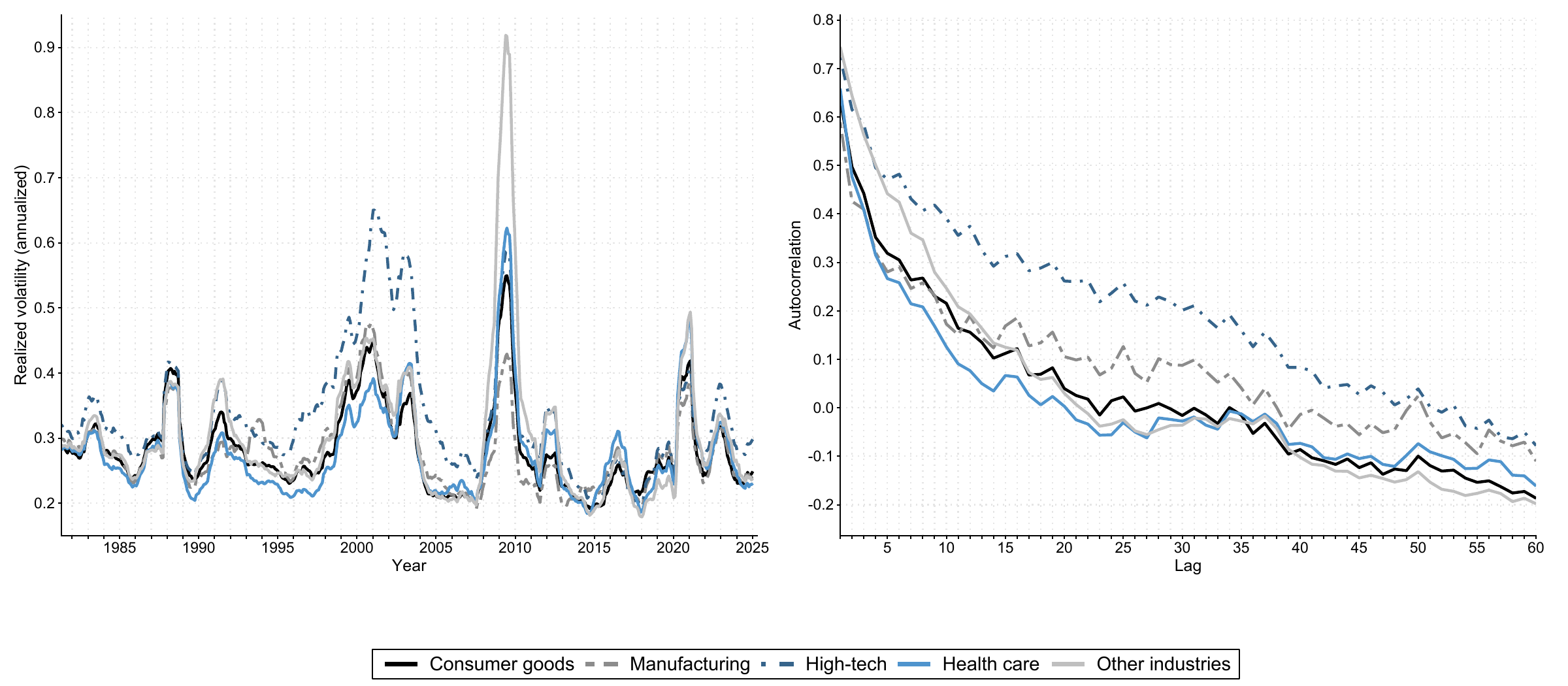}
	\caption{
		The left plot shows annualized monthly realized volatilities for
		S\&P~500 constituents, averaged by industry and smoothed using
		trailing 12-month moving averages. 
		We assign firms to sectors using the Standard Industrial Classification (SIC) codes and the five-industry classification available on Kenneth R. French’s data library \citep{FrenchData2026}. The right plot shows the autocorrelation functions of the corresponding unsmoothed monthly
		industry-average realized volatility series.
	}
	\alttext{Two line charts for five industry groups of S\&P 500 stocks: consumer
		goods, manufacturing, high-tech, health care and other industries. Left:
		smoothed annualized realized volatility from the early 1980s to 2025. All five
		series rise and fall together, with common spikes around 2000--2002, 2008--2009
		and 2020--2021; the largest spike occurs in 2008--2009. Right: autocorrelations up to
		lag 60. All series start between about 0.6 and 0.75 and decay slowly over many
		lags, with high-tech remaining the most persistent.}
	\label{fig:rv_acf_plot}
\end{figure}

To obtain a transparent first view of the cross-sectional and temporal dependence in realized volatility, we aggregate stocks into five industries. Figure~\ref{fig:rv_acf_plot} shows the resulting industry-average realized volatilities together with the autocorrelation functions of the corresponding (unsmoothed) time series.

The industry series exhibit substantial comovement, with volatility rising and falling together across sectors. Their autocorrelation functions also
decay slowly, indicating strong persistence across all five industries.
These patterns are consistent with the common variation and strong temporal
dependence in realized stock volatility documented by
\citet{Andersen2001}.
Despite these common patterns, the magnitude of volatility fluctuations and
the degree of persistence vary across industries. 

Figure~\ref{fig:corr_heatmaps} in \ref{app:figures} complements this
evidence at the firm level. Realized-volatility correlations are positive both
within and across industries, indicating that the broad comovement is not
confined to firms within the same sector. After removing the variation
captured by three principal components, cross-industry correlations become
small, while positive dependence remains within several industries. Taken together, the results point to a pervasive component in realized volatility alongside more localized within-industry dependence.

\subsection{Forecasting setup}
\label{Ch:VolatilityForecastingSetup}

For the forecasting analysis, we work with demeaned log realized volatility. With $RV_\tau$ denoting the vector of realized volatilities in month $\tau$, we estimate the pooled ARMA and factor specifications from
\begin{align}
	v_\tau
	&=
	\log(RV_\tau)-\hat\mu,
	&
	\hat\mu
	&=
	\frac{1}{\Tau}
	\sum_{\tau=1}^{\Tau}
	\log(RV_\tau),
	\label{eq:log_volatility}
\end{align}
where the logarithm and the mean are applied element-wise.
The resulting one-month-ahead forecast is transformed back to volatility
levels as
\begin{align}
	\hat\sigma_{\Tau+1\mid\Tau}
	&=
	\exp\!\left(
	\hat\mu+\hat v_{\Tau+1\mid\Tau}
	\right).
	\label{eq:logexp_fc}
\end{align}
The logarithmic transformation attenuates the influence of unusually large
volatility observations, while exponentiation maps the forecasts back to
positive volatility levels. We use the same log-exp transformation for all
specifications.\footnote{Since
	$\mathbb{E}[\exp(X)]\neq\exp(\mathbb{E}[X])$, a forecast that is optimal on
	the log scale need not remain optimal after exponentiation. However,
	\citet{Barigozzi2017,Barigozzi2020} show that this simple log-exp forecasting
	scheme performs well empirically and can outperform more sophisticated
	retransformation methods.}

\subsection{Parameter-pooled ARMA models}\label{Ch:Pooling}

The pooled ARMA approach exploits similarities in volatility dynamics by imposing common autoregressive and moving-average parameters within pre-specified groups of stocks. Relative to estimating a separate set of coefficients for each stock, pooling reduces the number of dynamic parameters and allows them to be estimated jointly from all volatility series within a group. 
Let $G$ denote the number of groups and $(n_1,\ldots,n_G)$ their respective sizes, with
\[
\sum_{g=1}^{G} n_g = N_t.
\]
The pooled ARMA$(p,p)$ model for demeaned log volatility is
\begin{align}\label{eq:pooled_arma}
	v_\tau
	=
	\sum_{\ell=1}^{p}
	A_\ell v_{\tau-\ell}
	+
	\sum_{\ell=1}^{p}
	B_\ell\varepsilon_{\tau-\ell}
	+
	\varepsilon_\tau,
\end{align}
where $\varepsilon_\tau$ collects the innovations of the individual stocks.
The coefficient matrices are block diagonal,
\begin{align}
	\begin{split}
	A_\ell
	&=
	\diag\!\left(
	a_{\ell,1}I_{n_1},
	\ldots,
	a_{\ell,G}I_{n_G}
	\right),
	\\
	B_\ell
	&=
	\diag\!\left(
	b_{\ell,1}I_{n_1},
	\ldots,
	b_{\ell,G}I_{n_G}
	\right),
	\qquad
	\ell=1,\ldots,p.
	\end{split}
\end{align}
Thus, stocks within the same group share the same ARMA coefficients, while
their innovations remain stock specific.

We consider three levels of pooling. Market-wide pooling ({ARMA(Mkt.)}) sets $G=1$ and imposes the same dynamic specification across the full cross-section. Industry pooling ({ARMA(Ind.)}) allows the coefficients to differ across the five industries used in Figure~\ref{fig:rv_acf_plot}. Finally, the univariate specification ({ARMA(Univ.)}) sets $G=N_t$ and estimates the dynamics separately for each stock, thereby removing the pooling restriction. The three specifications span a range from strong cross-sectional regularization to fully asset-specific dynamics, trading a smaller parameter dimension for greater restrictions on heterogeneity. In our main analysis, each pooling specification is implemented as an ARMA$(1,1)$ model; higher-order dynamics are considered in the robustness analysis.

\subsection{Factor models}
\label{Ch:FactorModels}

Factor models reduce the dimension of the volatility panel by separating variation shared across assets from variation specific to individual stocks. We consider both static and dynamic factor representations, which differ in how the factors affect individual volatilities over time.

Our SFM uses the principal-component representation
\begin{align}
	v_\tau
	&=
	\Lambda^{\mathrm{s}}F_\tau^{\mathrm{s}}
	+
	u_\tau,
	\label{eq:static_factor}
\end{align}
where $\Lambda^{\mathrm{s}}F_\tau^{\mathrm{s}}$ is the common component and
$u_\tau$ contains the remaining stock-specific variation. The factors
$F_\tau^{\mathrm{s}}$ and the loading matrix $\Lambda^{\mathrm{s}}$ are
estimated by principal components following \citet{Stock2002}. The number
of static factors is selected using the $PC_{p1}$ criterion of
\citet{Bai2002}, with the penalty calibrated according to the refinement of
\citet{Alessi2010}.\footnote{Figure~\ref{fig:nof_factors_plot} in \ref{app:figures} reports the numbers of static and dynamic factors selected across the rolling estimation windows.}

To obtain one-month-ahead forecasts, we follow the diffusion-index approach
of \citet{Stock2002a} and estimate a direct forecasting regression separately
for each stock. The baseline SFM forecast is
\begin{align}
	\hat v_{i,\Tau+1\mid\Tau}^{\mathrm{SFM}}
	&=
	\hat\beta_i^{\top}\hat F_\Tau^{\mathrm{s}}
	+
	\hat u_{i,\Tau+1\mid\Tau}^{\mathrm{ARMA}},
	\label{eq:sfm_forecast}
\end{align}
where $\hat\beta_i$ is estimated from a regression of $v_{i,\tau+1}$ on
$\hat F_\tau^{\mathrm{s}}$. The idiosyncratic component
$v_\tau-\hat\Lambda^{\mathrm{s}}\hat F_\tau^{\mathrm{s}}$ is forecast
separately for each stock using an ARMA$(1,1)$ model.

The GDFM of \citet{Forni2015,Forni2017} extends the static representation by allowing the  factors to affect individual volatilities through stock-specific dynamic loading filters,
\begin{align}
	v_{i,\tau}
	&=
	\lambda_i(L)f_\tau
	+
	u_{i,\tau},
	\qquad
	i=1,\ldots,N_t,
	\label{eq:dynamic_factor}
	\\
	\lambda_i(L)
	&=
	\sum_{\ell=0}^{\infty}
	\lambda_{i,\ell}L^\ell.
\end{align}
The common component $\lambda_i(L)f_\tau$ therefore incorporates the temporal dynamics of the factors directly into the factor representation. The resulting one-month-ahead GDFM forecast is
\begin{align}
	\hat v_{i,\Tau+1\mid\Tau}^{\mathrm{GDFM}}
	&=
	\sum_{\ell=1}^{p_f}
	\hat\lambda_{i,\ell}
	\hat f_{\Tau+1-\ell}
	+
	\hat u_{i,\Tau+1\mid\Tau}^{\mathrm{ARMA}}.
	\label{eq:gdfm_forecast}
\end{align}
In the empirical implementation, we truncate the one-sided loading filter
after $p_f=20$ lags, following \citet{Hoermann2021}.\footnote{The
	coefficients of the stable one-sided representation decay with the lag, and
	alternative truncation lengths yield nearly identical empirical results.} The number of
dynamic factors is selected using the $IC^{\Tau}_{2;N}$ criterion of
\citet{Hallin2007}. \ref{Ch:Appendix_FactorModel} provides details on the GDFM estimation and the construction of the forecasts.

\medskip
\noindent\textit{Remark (forecasting extensions).}
The baseline SFM uses only the most recent estimated factors in the direct
forecasting regression. Since contemporaneous static factors need not exhaust
dynamically common variation \citep{Barigozzi2026}, we additionally consider
SFM(lf), which includes one lag of the estimated factors. Following the
broader diffusion-index forecasting framework of \citet{Stock2002a}, we also
consider SFM(lrv), which augments the factor regression with the stock's own
most recent log realized volatility. Both extensions retain the same
principal-component factors but modify the forecasting regression and the
residual component subsequently forecast by ARMA.

\subsection{Parameter pooling and factor representations}
\label{Ch:ModelComparison}

The pooled ARMA, GDFM, and SFM achieve parsimony through different restrictions on the volatility process. The GDFM reduces the dimension of common variation through a small number of dynamic factors with stock-specific loading filters, whereas the SFM uses a low-dimensional contemporaneous factor representation and introduces additional dynamics at the forecasting stage. Parameter pooling takes a different approach. It retains the individual volatility series but reduces the number of dynamic parameters by imposing common autoregressive and moving-average coefficients within pre-specified groups. In fact, the pooled ARMA can be written as a factor model in which contemporaneous factor exposures remain stock specific, while the dynamic loadings are restricted by the pooling structure. This representation makes explicit where pooling gains parsimony and which forms of dynamic heterogeneity it rules out.

To see this, write Equation~\eqref{eq:pooled_arma} in lag-polynomial form as
\begin{align}
	A(L)v_\tau
	&=
	B(L)\varepsilon_\tau,
	\label{eq:pooled_arma_lag_form}
\end{align}
where
\begin{align}
	\begin{split}
		A(L)
		&=
		I-
		\sum_{\ell=1}^{p}
		A_\ell L^\ell
		=
		\operatorname{diag}\!\left(
		a_1(L)I_{n_1},
		\ldots,
		a_G(L)I_{n_G}
		\right),
		\\
		B(L)
		&=
		I+
		\sum_{\ell=1}^{p}
		B_\ell L^\ell
		=
		\operatorname{diag}\!\left(
		b_1(L)I_{n_1},
		\ldots,
		b_G(L)I_{n_G}
		\right).
	\end{split}
	\label{eq:pooled_arma_group_filters}
\end{align}
Thus, stocks within group $g$ share the ARMA polynomials $a_g(L)$ and
$b_g(L)$.

To relate the pooled ARMA to factor models, write its innovations as
\begin{align}
	\varepsilon_\tau
	&=
	\Lambda f_\tau+e_\tau,
	\label{eq:pooled_arma_innovation_factors}
\end{align}
where $f_\tau\in\mathbb R^q$ collects any common factors and $e_\tau$ the
remaining innovations. We allow $q\geq0$, with $q=0$ corresponding to the
case without a common component, so that $\varepsilon_\tau=e_\tau$. Since
$\varepsilon_\tau$ is white noise, any factor structure at the innovation
level is contemporaneous.

Under stability, the transfer function of the pooled ARMA is
\begin{align}
	H(L)
	&=
	A(L)^{-1}B(L)
	=
	\operatorname{diag}\!\left(
	h_1(L)I_{n_1},
	\ldots,
	h_G(L)I_{n_G}
	\right),
	\label{eq:pooled_arma_transfer_function}
\end{align}
where $h_g(L)=a_g(L)^{-1}b_g(L)$. Substituting
Equation~\eqref{eq:pooled_arma_innovation_factors} into
Equation~\eqref{eq:pooled_arma_lag_form} yields
\begin{align}
	v_\tau
	&=
	H(L)\Lambda f_\tau
	+
	H(L)e_\tau.
	\label{eq:pooled_arma_dynamic_factor}
\end{align}
The first term is the dynamically filtered common component of demeaned log
volatility; the second is its dynamically filtered idiosyncratic
component. Importantly, both are governed by the same group-specific transfer function.

Let $\lambda_i$ denote the loading vector of stock $i$. For a stock belonging
to group $g$, Equation~\eqref{eq:pooled_arma_dynamic_factor} becomes
\begin{align}
	v_{i,\tau}
	&=
	h_g(L)\lambda_i f_\tau
	+
	h_g(L)e_{i,\tau}.
	\label{eq:pooled_arma_group_factor}
\end{align}
The implied dynamic loading filter is therefore restricted to
\begin{align}
	\lambda_i(L)
	&=
	h_g(L)\lambda_i.
	\label{eq:pooled_arma_loading_restriction}
\end{align}
The contemporaneous loading vector $\lambda_i$ remains stock-specific, but stocks within the same group share the temporal response profile $h_g(L)$. Equivalently, the loading vector at each lag is a scalar multiple of $\lambda_i$, with the scalar determined by the group-specific filter. The GDFM does not impose this restriction and allows the dynamic loading filters to vary across stocks. The pooled ARMA further applies the same filter $h_g(L)$ to the idiosyncratic component, so common and idiosyncratic variation cannot follow different dynamics within a group. Under market-wide pooling these restrictions apply to the full cross-section, whereas industry pooling allows the filter to differ across industries.

The same representation also implies a static factor structure. Define the
group-specific filtered factors
\begin{align}
	\widetilde f_\tau^{(g)}
	&=
	h_g(L)f_\tau,
	\qquad
	g=1,\ldots,G,
\end{align}
and collect them in
\begin{align}
	\widetilde f_\tau
	&=
	\begin{pmatrix}
		\widetilde f_\tau^{(1)\top}
		&
		\cdots
		&
		\widetilde f_\tau^{(G)\top}
	\end{pmatrix}^{\top}.
\end{align}
Equation~\eqref{eq:pooled_arma_dynamic_factor} can then be written as
\begin{align}
	v_\tau
	&=
	\widetilde\Lambda\widetilde f_\tau
	+
	\widetilde e_\tau,
	\label{eq:pooled_arma_static_factor}
	\\
	\widetilde\Lambda
	&=
	\operatorname{diag}\!\left(
	\Lambda_1,\ldots,\Lambda_G
	\right),
	&
	\widetilde e_\tau
	&=
	H(L)e_\tau.
	\nonumber
\end{align}
This representation contains at most $Gq$ static factors. These are not
distinct primitive sources of common variation but group-specific filtered
versions of the same $q$ innovation factors. The corresponding loading
matrix is sparse because each factor block loads only on stocks belonging to
the associated group.

This exact static representation should be distinguished from the pervasive
factor structure estimated by principal components. Each filtered factor
$\widetilde f_\tau^{(g)}$ loads only on stocks in group $g$ and  can therefore be local rather than pervasive in the full cross-section, particularly when
the group is small. Such weak or local factors may not be recovered reliably
by principal components \citep{Boivin2006,Onatski2012}. More generally,
\citet{Barigozzi2026} emphasize that the dynamically common and statically
common components need not coincide. An SFM retaining only a
few pervasive factors can consequently omit some of the group-specific
dynamics represented by the pooled ARMA.

The converse distinction is equally important. The empirical SFM does not impose the group-specific dynamic restrictions implied by
pooling. Its forecasting coefficients are estimated separately for each
stock, while the remaining component is modeled separately. 
The pooled ARMA and the SFM therefore impose different
restrictions and are not nested within one another.

Taken together, these representations reveal the trade-offs underlying the
three approaches. Parameter pooling is most attractive when the temporal
dynamics are sufficiently similar within the chosen groups. Sharing a small
number of dynamic parameters across many stocks can then reduce estimation
uncertainty, irrespective of whether the innovations contain a common factor
component. This mechanism is consistent with \citet{Bollerslev_2018}, who
find that pooling volatility dynamics can improve out-of-sample forecasts
despite the restrictions it imposes. 
The GDFM allows stocks to respond differently to common variation, relaxing the homogeneity imposed by pooling at the cost of estimating more dynamic parameters. The SFM imposes a different form of dimension reduction: it
focuses on a small number of pervasive contemporaneous factors while allowing
stock-specific forecasting relationships and separate dynamics in the
remaining component. Its relative performance therefore depends on whether
the relevant common variation is adequately captured by this low-dimensional
factor space.

\section{Covariance Forecasting and Benchmarks}
\label{Ch:CovarianceForecasting}

To obtain forecasts of the full conditional covariance matrix, we combine one-month-ahead stock volatility forecasts with correlation matrix estimates that are identical across all previously discussed models.
Holding the correlation model fixed ensures that differences in forecast performance are attributable solely to differences in the volatility forecasts.

For comparison, we also consider covariance approaches that exploit commonality in returns or do not model volatility dynamics explicitly.

\subsection{Correlation and covariance forecasts}
\label{Ch:CorrelationForecasts}
We decompose the conditional covariance matrix as
\begin{align}
	\Sigma_\tau
	&=
	D_\tau R_\tau D_\tau,
	\label{eq:covariance_decomposition}
\end{align}
where $D_\tau$ contains the conditional return volatilities and $R_\tau$ denotes the conditional correlation matrix. The volatility models developed in Section~\ref{Ch:Methodology} determine the forecast of $D_\tau$, while $R_\tau$ is forecast using the correlation specification introduced below. The resulting one-month-ahead covariance forecast is
\begin{align}
	\widehat\Sigma_{\Tau+1\mid\Tau}
	&=
	\widehat D_{\Tau+1\mid\Tau}
	\widehat R_{\Tau+1\mid\Tau}
	\widehat D_{\Tau+1\mid\Tau},
	\label{eq:covariance_forecast}
	\\
	\widehat D_{\Tau+1\mid\Tau}
	&=
	\diag\!\left(
	\widehat\sigma_{1,\Tau+1\mid\Tau},
	\ldots,
	\widehat\sigma_{N_t,\Tau+1\mid\Tau}
	\right).
	\nonumber
\end{align}
For the correlation component, we build on the DCC framework of \citet{Engle2002}. Rather than updating correlations
from the outer product of standardized returns as in the standard DCC model,
we use realized correlation matrices following \citet{Bollerslev_2020} and \citet{Honig2025}.
These matrices summarize the information contained in the 21 daily returns observed within each month, which allows the correlation dynamics to be modeled at the same monthly frequency as the volatility	forecasts.

For month $\tau$, let $r_{\tau,k}$ denote the vector of daily stock returns on
day $k$. The corresponding realized covariance and correlation matrices
are
\begin{align}
	RCOV_\tau
	&=
	\sum_{k=1}^{21}
	r_{\tau,k}r_{\tau,k}^{\top},
	\label{eq:realized_covariance}
	\\
	RCOR_\tau
	&=
	\diag(RCOV_\tau)^{-1/2}
	RCOV_\tau
	\diag(RCOV_\tau)^{-1/2}.
	\label{eq:realized_correlation}
\end{align}
Using $RCOR_\tau$ as the realized measure of monthly dependence, the DCC
dynamics are specified as
\begin{align}
	Q_\tau
	&=
	\left(1-\alpha_q-\beta_q\right)C
	+
	\alpha_q RCOR_{\tau-1}
	+
	\beta_q Q_{\tau-1},
	\label{eq:rdcc_dynamics}
	\\
	R_\tau
	&=
	\diag(Q_\tau)^{-1/2}
	Q_\tau
	\diag(Q_\tau)^{-1/2}.
	\label{eq:rdcc_normalization}
\end{align}
The recursion evolves the positive definite matrix $Q_\tau$, without constraining its diagonal
elements to equal one. Rescaling by its diagonal produces the
corresponding conditional correlation matrix $R_\tau$.

The long-run correlation matrix $C$ is estimated from daily returns using
the nonlinear-shrinkage estimator of \citet{Ledoit2022}. 
We estimate the dynamic parameters \((\alpha_q,\beta_q)\) using the standard DCC likelihood applied to monthly returns divided by their realized volatility,
\begin{align}
	r^{(\mathrm{dev})}_{i,\tau}
	&=
	\frac{
		\sum_{k=1}^{21}r_{i,\tau,k}
	}{
		RV_{i,\tau}
	}.
	\label{eq:rdcc_devolatilized_returns}
\end{align}
 Following \citet{Engle2019} and \citet{Pakel2021}, we optimize the likelihood over adjacent asset pairs rather than jointly over the full cross-section. The resulting pairwise composite likelihood reduces the high-dimensional estimation problem to a sequence of bivariate likelihood evaluations and avoids repeated inversion of the full $N_t$-dimensional correlation matrix.

With $\widehat C$, $\widehat\alpha_q$, and $\widehat\beta_q$ in hand, the
one-month-ahead pseudo-correlation forecast is
\begin{align}
	\widehat Q_{\Tau+1\mid\Tau}
	&=
	\left(
	1-\widehat\alpha_q-\widehat\beta_q
	\right)\widehat C
	+
	\widehat\alpha_q RCOR_\Tau
	+
	\widehat\beta_q\widehat Q_\Tau.
	\label{eq:rdcc_forecast}
\end{align}
Normalizing $\widehat Q_{\Tau+1\mid\Tau}$ as in Equation~\eqref{eq:rdcc_normalization} yields the correlation forecast $\widehat R_{\Tau+1\mid\Tau}$. We refer to this specification as RDCC-NL.

\subsection{Benchmarks}
\label{Ch:CovarianceBenchmarks}
We compare the portfolio performance obtained using the covariance forecasts of the methods described in Sections~\ref{Ch:Methodology} and~\ref{Ch:CovarianceForecasting} with the performance of three covariance benchmarks and the equally weighted portfolio. The benchmarks include a standard conditional covariance estimator, a factor model approach that exploits commonality in returns rather than volatilities, and an unconditional covariance estimator.
As these methods are established in the literature, we summarize only the
features relevant for our empirical comparison and refer to the original
studies for further details.

\medskip
\noindent\textit{RDCC-NL.}
Following \citet{Honig2025}, monthly conditional volatilities are estimated using a realized \mbox{GARCH(1,1)} specification. The resulting volatility forecasts are combined with the RDCC-NL correlation forecasts described in Section~\ref{Ch:CorrelationForecasts}.

\medskip
\noindent\textit{AFM-RDCC-NL.}
We consider a monthly adaptation of the approximate factor covariance
estimator of \citet{DeNard2021}. Daily stock returns are decomposed into a
common component driven by a market factor and an idiosyncratic component. Monthly realized variances are computed for the market factor and the idiosyncratic components, and the conditional variances are each forecast using realized GARCH models. The common covariance forecast is constructed from the estimated loadings and the forecast factor variance, while the idiosyncratic covariance forecast combines the idiosyncratic variance forecasts with correlations obtained from RDCC-NL. The two covariance forecasts are subsequently recombined.

\medskip
\noindent\textit{NL.}
As an unconditional covariance benchmark, we use the nonlinear shrinkage
estimator of \citet{Ledoit2022}, applied to daily returns over each estimation
window. The estimate is scaled to a monthly frequency by multiplying by 21. This scaling does not affect minimum variance portfolio weights but facilitates comparisons between the estimates and their realized counterparts in Section~\ref{sec:stat_perf}.

\medskip
\noindent\textit{$1/N$.} 
The equally weighted portfolio. A widely used benchmark in empirical portfolio selection studies, which has been advocated by, for example, \citet{DeMiguel2009}. The approach is model-free, requires no parameter estimation, and can be conceptually understood as using the identity matrix as an estimate of the covariance matrix in portfolio selection.

\section{Empirical Analysis}\label{Ch:empirical_results}

\subsection{Data}\label{Ch:data}
We use data on the constituents of the S\&P 500 stock market index. Historical listings of the index constituents and data on daily total returns are obtained from LSEG Datastream. The return dataset covers the period from the beginning of January 1980 until the end of February 2025. Public holidays resulting in market closures affecting the trading of most stocks are excluded to ensure consistent trading sequences. 

In terms of data cleaning, we follow the same practices as in prior literature and employ rule-based filters to eliminate implausible return observations that, while rare, can distort measured volatility or the returns of selected portfolios.
As in the study by \citet{Landis2021}, we remove observations with excessive reversal effects, defined as returns exceeding 100\% ($-50\%$) followed by a return below $-50\%$ (100\%) on the stock’s next trading day.
Additionally, we follow \citet{Annaert2013} by excluding observations with likely decimal misreporting, specifically returns above 400\% or below $-85\%$. Removed return observations are replaced at the stock level using linear interpolation between the returns of the preceding and subsequent trading days. To mitigate the influence of illiquid stocks, we exclude stocks from the listings if their monthly realized volatility, as defined in Equation~\eqref{eq:RV}, is below 0.5\% in the month prior to their listing. This is sometimes observed among firms that remain listed during takeover or merger processes, when prices are effectively fixed around the offer value and trading activity is minimal.

\subsection{Estimation framework}\label{subsec:benchmark_methods}
We employ a rolling window approach to generate forecasts for all our models. At the end of each month, we estimate portfolio weights for all stocks using data from the previous ten years. After each iteration, the rolling window is advanced by one month. This implies that for methods that use monthly realized volatilities, estimation is performed on $\Tau = 120$ observations.

The stocks included in each estimation window are selected according to the most recent available index listing.
Given our rolling window design choice, we require stocks to have return histories spanning the entire in-sample period, as well as the subsequent $21$ days out-of-sample, which are used for backtesting. Along with the illiquidity filter discussed in Section~\ref{Ch:data}, this leads to the exclusion of some stocks. 
On average across all rolling windows, 0.054\% of stocks are removed owing to illiquidity, whereas 11.20\% are excluded because of incomplete return histories. Consequently, the number of stocks averages 445 and ranges from 411 to 470.

\subsection{Portfolio selection}\label{Ch:PortfolioSelection}
The covariance matrix estimates are evaluated based on the out-of-sample performance of their corresponding minimum variance portfolios. All covariance matrices used in portfolio selection are estimated at a monthly frequency, implying a one-month holding period. Portfolios are rebalanced as the rolling estimation window advances, with weights updated at the end of each month and subsequently evolving within the month as stock returns fluctuate. \newline
To assess portfolio performance, we consider the volatility of portfolio returns, measured by their standard deviation (SD), as well as average portfolio returns (AV) and Sharpe ratios (SR). All three measures are presented on an annualized basis and are computed from daily portfolio simple returns $r^{p}_{t,k}$ as follows:
\begin{align*}
	AV &= 
	\frac{12}{T}
	\sum_{t=1}^{T} \sum_{k=1}^{21} r^{p}_{t,k}, \\[1em]
	SD &= 
	\sqrt{
		\frac{12}{T}
		\sum_{t=1}^{T} \sum_{k=1}^{21}
		\left({r}^{p}_{t,k} - \bar{r}^{p}_{\vphantom{{t,k}}}  \right)^2
	}, \\[1em]
	SR &=
	\frac{
		\frac{12}{T}
		\sum_{t=1}^{T} \sum_{k=1}^{21}
		\left(r^{p}_{t,k} - r^{rf}_{t,k}\right)
	}{
		\sqrt{
			\frac{12}{T}
			\sum_{t=1}^{T} \sum_{k=1}^{21}
			\left(
			(r^{p}_{t,k} - r^{rf}_{t,k})
			-
			(\bar r^{p}_{\vphantom{{t,k}}} - \bar r^{rf}_{\vphantom{{t,k}}})
			\right)^2
		}
	},
\end{align*}
where $r^{rf}_{t,k}$ is the daily risk-free rate obtained from Kenneth R. French's data library \citep{FrenchData2026}, and $\bar{r}^{p}_{\vphantom{{t,k}}}$ ($\bar{r}^{rf}_{\vphantom{{t,k}}}$) represents the arithmetic mean of the portfolio returns (risk-free rate).
In our backtest analysis, we primarily base our argumentation on differences in the out-of-sample SD of portfolio returns, as it is the measure directly targeted by the minimum variance optimization. Expected returns are not included in the objective function and are therefore not directly optimized. From our perspective, this implies that the portfolio’s annualized return and, to a lesser extent, the SR arise as byproducts of the minimum variance portfolio and should mainly be interpreted as complementary metrics that provide additional information about portfolio performance. 

To assess the characteristics and practicability of the various portfolio strategies, we also consider average gross portfolio leverage (LEV), average fraction of negative portfolio weights (NEG), and the average monthly portfolio turnover rate (TO). These quantities are defined as
\begin{align*}
	\text{LEV} 
	&= \frac{1}{T}
	\sum_{t=1}^{T}
	\sum_{i=1}^{N_t} \left| \hat{w}_{i,t} \right|, \\
	\text{NEG} 
	&= \frac{1}{T}
	\sum_{t=1}^{T}
	\frac{1}{N_t}
	\sum_{i=1}^{N_t}
	\mathbbm{1}_{\{\hat{w}_{i,t} < 0\}}, \label{eq:portfolio_measures} \\
	\text{TO} 
	&= \frac{1}{T-1}
	\sum_{t=2}^{T}
	\frac{1}{2}
	\sum_{i=1}^{N_t}
	\left| \hat{w}_{i,t} - \hat{w}^{eom}_{i,t-1} \right|.
\end{align*}
Here, $\hat{w}_{i,t}$ denotes the portfolio weight of asset $i$ at month $t$. The vector $\hat{w}^{eom}_{t-1}$ represents the portfolio weights at the end of month $t-1$ obtained by passively holding the portfolio with weights $\hat{w}_{t-1}$ over that month. This adjustment also accounts for firms entering or exiting the investment universe at $t$, as entering (exiting) firms receive a portfolio weight of zero in $\hat{w}^{\mathrm{eom}}_{t-1}$ ($\hat{w}_t$). In general, portfolio turnover measures the extent to which portfolio holdings are reallocated, or ``turned over,'' when the portfolio is rebalanced. In non-leveraged portfolios, the maximum turnover is one, whereas turnover may exceed one when short selling is allowed. Although evaluating the portfolio performance net of transaction costs is beyond the scope of this study, the reported average turnover rate provides a rough indication of the transaction costs that would be incurred when rebalancing the portfolios in practice. 
Monthly transaction costs can be approximated by the product of portfolio turnover and the per-trade transaction cost, multiplied by two to account for both the buy and sell sides of each transaction.

\subsubsection{Global minimum variance portfolios}\label{subsec:gmv}
The GMV portfolio is obtained by solving
\begin{equation}
	\min_{w_t} \; w_{t}^\top \Sigma_{t} w_{t} 
	\quad \text{subject to} \quad w_{t}^\top \mathbf{1} = 1,
	\label{eq:min_var_unr}
\end{equation}
where $w_t = (w_{1,t}, \ldots, w_{N_{t}, t})^\top$ denotes the portfolio weights and 
$\mathbf{1} = (1, \ldots, 1)^\top$ is the $N_t$-dimensional vector of ones.
The analytical solution to \eqref{eq:min_var_unr} is
\begin{equation}
	w^{\mathrm{opt}}_{t}
	= \frac{\Sigma^{-1}_{t} \mathbf{1}}{\mathbf{1}^\top \Sigma^{-1}_{t} \mathbf{1}}.
	\label{eq:mv_solution}
\end{equation}
In practice, the covariance matrix \( \Sigma_{t} \) is generally unknown and therefore replaced by its estimate \( \hat{\Sigma}_{t} \), which yields the estimated optimal weights 
\begin{equation}
	\hat{w}^{\mathrm{opt}}_{t} 
	= \frac{\hat{\Sigma}^{-1}_{t} \mathbf{1}}{\mathbf{1}^\top \hat{\Sigma}^{-1}_{t} \mathbf{1}}.
	\label{eq:mv_solution_est}
\end{equation}
Table~\ref{tab:perf_gmv} shows the out-of-sample performance and portfolio characteristics of the GMV portfolios constructed using the different approaches presented in Sections~\ref{Ch:Methodology} and \ref{Ch:CovarianceForecasting}. The 1/N portfolio is consistently outperformed by all other portfolios, indicating the benefits of using statistical models in this empirical setting. Moreover, all portfolios based on conditional covariance matrix estimates produce portfolios with lower return volatility than the unconditional estimate of NL. By emphasizing recency, conditional covariance matrices produce portfolios that better reflect current market conditions, which unconditional estimates struggle to capture because they weight all temporal information uniformly.
\begin{table}[!h]
	\centering
	\scriptsize
	\caption{Performance measures for the GMV portfolios.}
	\label{tab:perf_gmv}
	\resizebox{\textwidth}{!}{%
		\setlength{\tabcolsep}{5pt}
		\begin{tabular*}{\textwidth}{l@{\hspace{8pt}}@{\extracolsep{\fill}}rrrrrrrr}
			\toprule
			\multicolumn{9}{c}{Evaluation period: January 1990 -- February 2025} \\
			\toprule
			Method 				& AV & SD & SR & MAX & MIN & NEG & LEV & TO \\
			\midrule
			1/N 				& 12.61 & 18.56 & 0.54 & 0.00\rlap{$^{\dagger}$} & 0.00\rlap{$^{\dagger}$} & 0.00 & 1.00 & 0.04 \\
			NL 					& 9.30 & 11.44 & 0.58 & 0.06 & -0.04 & 0.45 & 4.45 & 0.29 \\
			RDCC-NL	 	& 11.13 &10.34   & 0.82 &  0.09&  -0.04 &  0.48 & 4.69 & 1.66 \\
			AFM-RDCC-NL	 		& 9.51 & 11.24 & 0.61 & 0.10 & -0.06 & 0.46 & 4.96 & 0.77 \\
			
			SFM-RDCC-NL 		& 10.39 & 10.09 & 0.77 & 0.17 & -0.04 & 0.49 & 4.69 & 1.64 \\
			SFM(lrv)-RDCC-NL 	& 10.35 & 10.20 & 0.76 & 0.18 & -0.04 & 0.49 & 4.52 & 1.83 \\
			SFM(lf)-RDCC-NL 	& 10.16 & 10.55 & 0.71 & 0.16 & -0.04 & 0.49 & 4.55 & 1.91 \\
			GDFM-RDCC-NL 		& 9.97 & 9.87 & 0.74 & 0.19 & -0.04 & 0.49 & 4.68 & 1.53 \\
			
			ARMA(Univ.)-RDCC-NL & 11.01 & 9.78 & 0.86 & 0.19 & -0.04 & 0.49 & 4.43 & 1.72 \\
			ARMA(Ind.)-RDCC-NL  & 10.69 & 9.27 & 0.87 & 0.17 & -0.04 & 0.49 & 4.59 & 1.61 \\
			ARMA(Mkt.)-RDCC-NL  & 10.80 & \textbf{9.13\rlap{\textsuperscript{***}}} & 0.89 & 0.17 & -0.04 & 0.49 & 4.55 & 1.58 \\
			\bottomrule
		\end{tabular*}
	} 
	\begin{tablenotes}[flushleft]
		\scriptsize
		\item Reported annualized performance measures are the average portfolio return (AV), standard deviation (SD), and Sharpe ratio (SR). We further report the average maximum and minimum weights (MAX, MIN), fraction of negative weights (NEG), gross leverage (LEV), and monthly turnover (TO). All measures are calculated over the 8,841 out-of-sample simple daily returns of the corresponding portfolio. 
		Bold values identify the lowest SD. Asterisks (*, **, ***) indicate statistically significant differences in SD between the approach shown in bold font (best-performing) and each of the remaining approaches; the largest $p$-value across these pairwise tests is used to determine significance at the 10\%, 5\%, and 1\% levels, respectively. $^{\dagger}$ indicates positive values smaller than 0.005.
	\end{tablenotes}
\end{table}
\FloatBarrier
Pooled ARMA$(1,1)$ specifications delivered the lowest portfolio SDs among all approaches considered. In particular, ARMA(Mkt.)-RDCC-NL outperformed every other approach, with all pairwise differences in portfolio SD being statistically significant at the 1\% level.\footnote{Differences in SDs are tested using two-sided $p$-values computed with the prewhitened HAC method of \citet{Ledoit2011}.}

All constructed portfolios exhibit similarly high levels of leverage and, for the most part, a comparably large dispersion between minimum and maximum portfolio weights. This outcome can be explained by the well-known tendency of the GMV portfolio to generate extreme weights through the exploitation of offsetting positions among correlated assets \citep{Green1992}. In particular, the GMV strategy takes large offsetting positions in similar assets whose combined exposure remains low in volatility, which in turn can lead to high leverage and substantial dispersion in the minimum and maximum weights.
Turnover is higher for the portfolios that are based on estimated conditional covariance matrices than for the portfolio that uses the unconditional NL estimate. 
This reflects the time variation in conditional covariances, which leads to larger portfolio weight adjustments at each rebalancing date. Monthly turnover is similar across portfolios based on conditional estimates, except for the AFM-RDCC-NL portfolio, which has a lower turnover of 77\%. 

\subsubsection{Robustness checks}\label{Ch:Robustness}
\paragraph{\textit{Long-only minimum variance portfolios}}
To assess a setting in which instabilities arising from offsetting positions are mitigated, we follow \citet{Green1992} and evaluate the performance of minimum variance portfolios with an additional non-negativity constraint on portfolio weights:
\begin{equation}
	\begin{aligned}
		\min_{w_{t}} \quad & w_{t}^\top \Sigma_{t} w_{t} \\
		\text{subject to} \quad 
		& w_{t}^\top \mathbf{1} = 1, \\
		& w_{t} \ge 0.
	\end{aligned}
	\label{eq:min_var_long_only}
\end{equation}
Portfolio results for the constrained optimization setting are reported in Table~\ref{tab:perf_longonly}. 
\begin{table}[!h]
	\centering
	\scriptsize
	\caption{Performance measures for the restricted (long-only) portfolios.}
	\label{tab:perf_longonly}
	\resizebox{\textwidth}{!}{%
		\begin{tabular*}{\textwidth}{l@{\hspace{8pt}}@{\extracolsep{\fill}}rrrrrrrr}
			\toprule
			\multicolumn{9}{c}{Evaluation period: January 1990 -- February 2025} \\
			\toprule
			Method  		 & AV & SD & SR & MAX & MIN & NEG & LEV & TO \\
			\midrule
			1/N 			 & 12.61 & 18.56 & 0.54 & 0.00\rlap{$^{\dagger}$} & 0.00\rlap{$^{\dagger}$} & 0.00 & 1.00 & 0.04 \\
			NL 				 & 10.27 & 13.08 & 0.58 & 0.09 & 0.00\rlap{$^{\dagger}$} & 0.00 & 1.00 & 0.05 \\
			RDCC-NL   & 9.64 & 10.89 & 0.64 & 0.16 & 0.00\rlap{$^{\dagger}$} & 0.00 & 1.00 & 0.40 \\
			AFM-RDCC-NL  & 10.07 & 12.61 & 0.59 & 0.13 & 0.00\rlap{$^{\dagger}$} & 0.00 & 1.00 & 0.22 \\
			SFM-RDCC-NL 	 & 9.46 & 10.34 & 0.66 & 0.24 & 0.00\rlap{$^{\dagger}$} & 0.00 & 1.00 & 0.45 \\
			SFM(lrv)-RDCC-NL & 9.29 & 10.39 & 0.64 & 0.25 & 0.00\rlap{$^{\dagger}$} & 0.00 & 1.00 & 0.51 \\
			SFM(lf)-RDCC-NL  & 9.25 & 10.59 & 0.62 & 0.24 & 0.00\rlap{$^{\dagger}$} & 0.00 & 1.00 & 0.52 \\
			GDFM-RDCC-NL 	 & 8.87 & 10.18 & 0.61 & 0.26 & 0.00\rlap{$^{\dagger}$} & 0.00 & 1.00 & 0.43 \\
			ARMA(Univ.)-RDCC-NL & 8.99 & 10.15 & 0.62 & 0.27 & 0.00\rlap{$^{\dagger}$} & 0.00 & 1.00 & 0.48 \\
			ARMA(Ind.)-RDCC-NL  & 8.94 & 9.68 & 0.65 & 0.24 & 0.00\rlap{$^{\dagger}$} & 0.00 & 1.00 & 0.43 \\
			ARMA(Mkt.)-RDCC-NL  & 9.05 & \textbf{9.60\rlap{\textsuperscript{**}}} & 0.67 & 0.25 & 0.00\rlap{$^{\dagger}$} & 0.00 & 1.00 & 0.43 \\
			\bottomrule
		\end{tabular*}
	}
	\begin{tablenotes}[flushleft]
		\scriptsize
		\item Reported annualized performance measures are the average   portfolio return (AV), standard deviation (SD), and Sharpe ratio (SR). We further report the average maximum and minimum weights (MAX, MIN), fraction of negative weights (NEG), gross leverage (LEV), and monthly turnover (TO). All measures are calculated over the 8,841 out-of-sample simple daily returns of the corresponding portfolio. 
		Bold values identify the lowest SD. Asterisks (*, **, ***) indicate statistically significant differences in SD between the approach shown in bold font (best-performing) and each of the remaining approaches; the largest $p$-value across these pairwise tests is used to determine significance at the 10\%, 5\%, and 1\% levels, respectively. $^{\dagger}$ indicates positive values smaller than 0.005.
	\end{tablenotes}
\end{table}
\FloatBarrier
The relative ranking of portfolios in terms of SD is nearly identical to that of the unconstrained GMV case, with the pooled ARMA approaches continuing to perform best. As expected, the long-only constraint uniformly raises portfolio SDs, as it limits the formation of variance-reducing hedging positions. The reduced flexibility of the optimization problem also leads to smaller performance differentials among the volatility models introduced in Section~\ref{Ch:Methodology}. 
Nevertheless, the portfolio SD of the ARMA(Mkt.)-RDCC-NL model is 8 basis points lower than that of the industry-pooled variant, a difference that is statistically significant at the 5\% level. 

The non-negativity constraint on portfolio weights, in combination with the budget restriction, indirectly limits portfolio gross leverage. Compared with the unconstrained portfolios, this results in a substantial reduction in portfolio turnover across all estimation approaches.

\paragraph{Leverage- and turnover-constrained portfolios}
In practice, portfolio managers often operate under limits on leverage and short selling given the additional risks posed by both practices \citep{BrunnermeierPedersen2009, Engelberg2018}. 
They may also find it desirable to control turnover, as large adjustments to portfolio positions during rebalancing often entail substantial transaction costs. \newline 
To reflect these practical considerations, we evaluate a minimum variance portfolio on which we impose the corresponding constraints. Specifically, we set the maximum gross leverage $\mathscr{L}=1.6$, monthly turnover limit $\mathscr{T}=0.3$, minimum asset weight $w_{\min}=-0.05$, and maximum asset weight \mbox{$w_{\max}=0.05$}.
\begin{equation}
\begin{aligned}
	\min_{w_t} \quad
	& w_t^\top \Sigma_t w_t \\[2pt]
	\text{subject to} \quad
	& w_t^\top \mathbf{1}_{N_t}=1, \qquad 
	w_{i,t}\in[w_{\min},w_{\max}] \\
	& \sum_{i=1}^{N_t}|w_{i,t}|
	\leq \mathscr{L},\qquad
	\frac{1}{2}
	\sum_{i=1}^{N_t}
	\left|w_{i,t}-w^{\mathrm{eom}}_{i,t-1}\right|
	\leq \mathscr{T}.
\end{aligned}
\end{equation}
The corresponding portfolio results are reported in Table~\ref{tab:perf_13030_turnover}. 
Together with the budget constraint, the leverage constraint permits a long exposure of up to 130\% and a short exposure of up to 30\%, corresponding to the widely used ``130/30'' portfolio structure \citep{LoPatel2008}.
\begin{table}[!h]
	\centering
	\scriptsize
	\caption{Performance measures for the leverage- and turnover-constrained 130/30 portfolios.}
	\label{tab:perf_13030_turnover}
	\resizebox{\textwidth}{!}{%
		\begin{tabular*}{\textwidth}{l@{\hspace{8pt}}@{\extracolsep{\fill}}rrrrrrrr}
			\toprule
			\multicolumn{9}{c}{Evaluation period: January 1990 -- February 2025} \\
			\toprule
			Method & AV & SD & SR & MAX & MIN & NEG & LEV & TO \\
			\midrule
			1/N
			& 12.61 & 18.56 & 0.54
			& 0.00\rlap{$^{\dagger}$}
			& 0.00\rlap{$^{\dagger}$}
			& 0.00 & 1.00 & 0.04 \\
			
			NL
			& 9.13 & 12.01 & 0.54
			& 0.05 & -0.04 & 0.35 & 1.60 & 0.09 \\
			
			RDCC-NL
			& 9.24 & 10.65 & 0.62
			& 0.05 & -0.03 & 0.34 & 1.60 & 0.30 \\
			
			AFM-RDCC-NL
			& 9.31 & 11.56 & 0.58
			& 0.05 & -0.04 & 0.34 & 1.60 & 0.26 \\
			
			SFM-RDCC-NL
			& 9.58 & 10.77 & 0.64
			& 0.05 & -0.03 & 0.37 & 1.60 & 0.30 \\
			
			SFM(lrv)-RDCC-NL
			& 9.58 & 10.82 & 0.64
			& 0.05 & -0.03 & 0.37 & 1.60 & 0.30 \\
			
			SFM(lf)-RDCC-NL
			& 9.66 & 10.80 & 0.65
			& 0.05 & -0.03 & 0.37 & 1.60 & 0.30 \\
			
			GDFM-RDCC-NL
			& 9.32 & 10.60 & 0.63
			& 0.05 & -0.03 & 0.37 & 1.60 & 0.30 \\
			
			ARMA(Univ.)-RDCC-NL
			& 9.29 & 10.56 & 0.63
			& 0.05 & -0.03 & 0.37 & 1.60 & 0.30 \\
			
			ARMA(Ind.)-RDCC-NL
			& 9.41 & \textbf{10.20} & 0.66
			& 0.05 & -0.03 & 0.37 & 1.60 & 0.30 \\
			
			ARMA(Mkt.)-RDCC-NL
			& 9.47 & 10.22 & 0.67
			& 0.05 & -0.03 & 0.37 & 1.60 & 0.30 \\
			\bottomrule
		\end{tabular*}
	}
	\begin{tablenotes}[flushleft]
		\scriptsize
		\item Reported annualized performance measures are the average portfolio return
		(AV), standard deviation (SD), and Sharpe ratio (SR). We further report
		the average maximum and minimum weights (MAX, MIN), fraction of negative
		weights (NEG), gross leverage (LEV), and monthly turnover (TO). All
		measures are calculated over the 8,841 out-of-sample simple daily returns
		of the corresponding portfolio. Bold values identify the lowest SD.
		$^{\dagger}$ indicates positive values smaller than 0.005.
	\end{tablenotes}
\end{table}
\FloatBarrier
The combination of these constraints reduces flexibility in the optimization and thus limits the extent to which differences in covariance estimates translate into distinct portfolio weights. Consequently, portfolio SDs are more closely clustered than in the unconstrained setting.

ARMA(Ind.)-RDCC-NL achieves the lowest annualized SD of 10.20\%, only two basis points below the 10.22\% obtained by ARMA(Mkt.)-RDCC-NL. Importantly, however, the portfolios constructed using ARMA(Ind.)-RDCC-NL and ARMA(Mkt.)-RDCC-NL each have a lower SD than those constructed using any of the remaining approaches, with every such difference significant at the 1\% level.

\paragraph{\textit{Subperiod analysis}}
We assess whether the results of the GMV portfolios are stable over time or are driven by performance differences in specific subperiods. The daily portfolio return series are partitioned into non-overlapping five-year periods. Table~\ref{tab:window_sd} shows that the ARMA(Mkt.)-RDCC-NL achieved the best performance in all the subperiods, tying with the GDFM-RDCC-NL in subperiod ``1990--1994'' and with the ARMA(Ind.)-RDCC-NL in ``2000--2004''.
\begin{table}[htb]
	\centering
	\scriptsize
	\caption{Annualized portfolio standard deviations over non-overlapping 5-year subperiods.}
	\resizebox{\textwidth}{!}{%
		\setlength{\tabcolsep}{4pt}
		\begin{tabular*}{\textwidth}{l@{\hspace{2pt}}@{\extracolsep{\fill}}rrrrrrr}
			\toprule		
			& \multicolumn{7}{c}{SD} \\
			\cmidrule(lr){2-8}
			Method & \shortstack{\scriptsize 1990\\\scriptsize --1994}
			& \shortstack{\scriptsize 1995\\\scriptsize --1999}
			& \shortstack{\scriptsize 2000\\\scriptsize --2004}
			& \shortstack{\scriptsize 2005\\\scriptsize --2009}
			& \shortstack{\scriptsize 2010\\\scriptsize --2014}
			& \shortstack{\scriptsize 2015\\\scriptsize --2019}
			& \shortstack{\scriptsize 2020\\\scriptsize --2025} \\
			\midrule
			1/N & 12.14 & 13.56 & 19.16 & 27.05 & 17.61 & 13.47 & 22.11 \\ 
			NL &  8.55 &  9.83 & 11.99 & 14.54 &  9.46 &  9.90 & 14.22 \\ 
			RDCC-NL &  7.62 &  9.05 & 10.80 & 11.69 &  8.76 &  9.19 & 13.87 \\ 
			AFM-RDCC-NL &  8.22 &  9.58 & 11.51 & 13.76 &  9.21 &  9.39 & 15.11 \\ 
			SFM-RDCC-NL &  7.59 &  9.16 & 10.43 & 11.48 &  8.08 &  8.96 & 13.54  \\ 
			SFM(lrv)-RDCC-NL &  7.62 &  9.37 & 10.55 & 11.67 &  8.43 &  8.90 & 13.55 \\ 
			SFM(lf)-RDCC-NL &  7.70 &  9.28 & 10.68 & 12.03 &  8.64 &  9.09 & 14.69 \\ 
			GDFM-RDCC-NL &  \textbf{7.46} &  8.85 & 10.21 & 11.55 &  7.64 &  8.40 & 13.62 \\
			ARMA(Univ.)-RDCC-NL &  7.48 &  8.81 & 10.24 & 10.53 &  7.44 &  8.18 & 13.99 \\ 
			ARMA(Ind.)-RDCC-NL &  7.47 &  8.67 &  \textbf{9.83} &  9.26 &  7.44 &  8.23 & 12.80 \\ 
			ARMA(Mkt.)-RDCC-NL &  \textbf{7.46} &  \textbf{8.64} &  \textbf{9.83} & \textbf{8.98\rlap{\textsuperscript{***}}} &  \textbf{7.26\rlap{\textsuperscript{***}}} &  \textbf{8.04\rlap{\textsuperscript{***}}} & \textbf{12.54} \\  
			\bottomrule
		\end{tabular*}
	}
	\begin{tablenotes}[flushleft]
		\scriptsize
		\item Reported are the annualized SDs computed over non-overlapping five-year subperiods. Each subperiod begins with the earliest available observation in the starting year and includes all observations through the end of the final year. For the final subperiod (``2020--2025''), observations through February 2025 are included. Bold values identify the lowest SD. Asterisks (*, **, ***) indicate statistically significant differences in SD between the approach shown in bold font (best-performing) and each of the remaining approaches; the largest $p$-value across these pairwise tests is used to determine significance at the 10\%, 5\%, and 1\% levels, respectively.
	\end{tablenotes}
	\label{tab:window_sd}
\end{table}
\FloatBarrier
Three of the considered subperiods coincide with major episodes of economic stress.
Specifically, the ``2000--2004'' subperiod includes the dot-com collapse; the ``2005--2009'' phase coincides with the global financial crisis; and the COVID-19 pandemic is included in the ``2020--2025'' phase. In these phases, the ARMA(Mkt.)-RDCC-NL achieves substantial risk reduction compared with the best-performing non-pooling alternative, with portfolio SD lower by 3.7\% (``2000--2004''),  14.7\% (``2005--2009''), and 7.4\% (``2020--2025'').

\paragraph{\textit{Longer estimation window}}
As an additional robustness check, we assess the sensitivity of our results to an alternative rolling-window length. We extend the window from 10 to 15 years, corresponding to estimation samples of 180 (3,780) monthly (daily) observations. As shown in Table~\ref{tab:perf_gmv_15}, the overall level of portfolio SDs increases relative to the values shown in Table~\ref{tab:perf_gmv}. 
\begin{table}[htb]
	\centering
	\scriptsize
	\caption{Performance measures for the GMV portfolios (15-year rolling windows).}
	\label{tab:perf_gmv_15}
	\resizebox{\textwidth}{!}{%
		\setlength{\tabcolsep}{4pt}
		\begin{tabular*}{\textwidth}{l@{\hspace{8pt}}@{\extracolsep{\fill}}rrrrrrrr}
			\toprule
			\multicolumn{9}{c}{Evaluation period: January 1995 -- February 2025} \\
			\toprule
			Method 				& AV & SD & SR & MAX & MIN & NEG & LEV & TO \\
			\midrule
			1/N 				& 13.15 & 19.16 & 0.57 & 0.00\rlap{$^{\dagger}$} & 0.00\rlap{$^{\dagger}$} & 0.00 & 1.00 & 0.03 \\
			NL 					& 10.05 & 12.34 & 0.63 & 0.07 & -0.04 & 0.45 & 4.16 & 0.23 \\
			RDCC-NL 		& 10.95 & 10.82 & 0.80 & 0.11 & -0.05 & 0.48 & 4.91 & 1.72 \\
			AFM-RDCC-NL	 		& 9.53 & 11.99 & 0.60 & 0.11 & -0.06 & 0.46 & 4.95 & 0.78 \\
			
			SFM-RDCC-NL 		& 10.27 & 11.11 & 0.72 & 0.19 & -0.05 & 0.49 & 4.98 & 1.69 \\
			SFM(lrv)-RDCC-NL 	& 10.49 & 11.11 & 0.74 & 0.20 & -0.05 & 0.49 & 4.78 & 1.95 \\
			SFM(lf)-RDCC-NL 	& 10.13 & 11.43 & 0.68 & 0.18 & -0.05 & 0.49 & 4.86 & 1.99 \\
			GDFM-RDCC-NL 		& 9.78 & 10.74 & 0.70 & 0.22 & -0.05 & 0.49 & 4.88 & 1.56 \\
			
			ARMA(Univ.)-RDCC-NL & 10.24 & 10.35 & 0.77 & 0.22 & -0.04 & 0.49 & 4.59 & 1.79 \\
			ARMA(Ind.)-RDCC-NL  & 9.99 & 10.06 & 0.76 & 0.20 & -0.05 & 0.49 & 4.73 & 1.71 \\
			ARMA(Mkt.)-RDCC-NL  & 9.91 & \textbf{9.94\rlap{\textsuperscript{***}}} & 0.76 & 0.21 & -0.04 & 0.49 & 4.69 & 1.69 \\
			\bottomrule
		\end{tabular*}
	}
	\begin{tablenotes}[flushleft]
		\scriptsize
		\item Reported annualized performance measures are the average   portfolio return (AV), standard deviation (SD), and Sharpe ratio (SR). We further report the average maximum and minimum weights (MAX, MIN), fraction of negative weights (NEG), gross leverage (LEV), and monthly turnover (TO). All measures are calculated over the 7,581 out-of-sample simple daily returns of the corresponding portfolio. 
		Bold values identify the lowest SD. Asterisks (*, **, ***) indicate statistically significant differences in SD between the approach shown in bold font (best-performing) and each of the remaining approaches; the largest $p$-value across these pairwise tests is used to determine significance at the 10\%, 5\%, and 1\% levels, respectively. $^{\dagger}$ indicates positive values smaller than 0.005.
	\end{tablenotes}
\end{table}
\FloatBarrier
We explain this result by the later start of the out-of-sample period, which excludes the relatively low-volatility phase ``1990--1994'' reported in Table~\ref{tab:window_sd}.
Nevertheless, the ARMA(Mkt.)-RDCC-NL portfolio continues to exhibit a significantly lower SD than all other portfolios. The relative magnitude of the differences in model performances is comparable to those reported in Table~\ref{tab:perf_gmv}. The ARMA(Mkt.)-RDCC-NL portfolio has a 12 basis points lower SD than the ARMA(Ind.)-RDCC-NL (versus 14 basis points under a 10-year rolling window).
ARMA(Mkt.)-RDCC-NL outperforms the best SFM setup by 117 basis points and GDFM-RDCC-NL by 80 basis points (96 and 74 basis points, respectively, under a 10-year rolling window).

\paragraph{\textit{Different correlation model}}
Given that our evaluation centers on differences in the estimation of realized volatilities, we examine how the results change when the correlation model is replaced. In all approaches, we replace the monthly RDCC-NL estimates with DCC-NL correlation forecasts, following \citet{Engle2019}. The DCC-NL model is estimated at the daily frequency, and its forecasts are aggregated over 21 trading days to obtain monthly correlation matrices following \citet{DeNard2021}. Table~\ref{tab:perf_gmv_dcc} shows that the performance of the market-pooled ARMA approach remains largely unchanged by this modification. The SFM, GDFM, and remaining ARMA approaches exhibit modest performance improvements. Apart from the AFM-DCC-NL, a further notable feature is the reduction in portfolio turnover for approaches that model the conditional covariance matrix.
This reduction may be attributable to the DCC-NL method's reliance on multi-step forecasts, which tend to converge toward the unconditional correlation matrix. Therefore, correlations tend to vary less across rolling window estimations than under the monthly RDCC-NL forecasts, resulting in smaller position adjustments on portfolio rebalancing. 
\begin{table}[htb]
	\centering
	\scriptsize
	\caption{Performance measures for the GMV portfolios (DCC-NL correlation forecasts).}
	\label{tab:perf_gmv_dcc}
	\resizebox{\textwidth}{!}{%
		\setlength{\tabcolsep}{4pt}
		\begin{tabular*}{\textwidth}{l@{\hspace{8pt}}@{\extracolsep{\fill}}rrrrrrrr}
			\toprule
			\multicolumn{9}{c}{Evaluation period: January 1990 -- February 2025} \\
			\toprule
			Method 				& AV & SD & SR & MAX & MIN & NEG & LEV & TO \\
			\midrule
			1/N 				& 12.61 & 18.56 & 0.54 & 0.00\rlap{$^{\dagger}$} & 0.00\rlap{$^{\dagger}$} & 0.00 & 1.00 & 0.04 \\
			NL 					& 9.30 & 11.44 & 0.58 & 0.06 & -0.04 & 0.45 & 4.45 & 0.29 \\
			DCC-NL 		& 10.82 & 10.98 & 0.75 & 0.12 & -0.04 & 0.49 & 4.26 & 1.56 \\
			AFM-DCC-NL	 		& 9.21 & 10.75 & 0.61 & 0.10 & -0.05 & 0.46 & 4.45 & 0.97 \\
			
			SFM-DCC-NL 		    & 10.25 & 9.95 & 0.77 & 0.15 & -0.04 & 0.50 & 4.19 & 1.41 \\
			SFM(lrv)-DCC-NL 	& 10.29 & 10.01 & 0.76 & 0.17 & -0.04 & 0.50 & 4.07 & 1.56 \\
			SFM(lf)-DCC-NL 	    & 10.08 & 10.38 & 0.72 & 0.15 & -0.04 & 0.50 & 4.08 & 1.66 \\
			GDFM-DCC-NL 		&  10.14 & 9.74& 0.77 & 0.18 & -0.04 & 0.49 & 4.20 & 1.26 \\
			
			ARMA(Univ.)-DCC-NL  & 10.87 & 9.62 & 0.86 & 0.18 & -0.04 & 0.50 & 4.07 & 1.41 \\
			ARMA(Ind.)-DCC-NL   & 10.41 & 9.22 & 0.84 & 0.16 & -0.04 & 0.49 & 4.22 & 1.28 \\
			ARMA(Mkt.)-DCC-NL   & 10.45 & \textbf{9.13\rlap{\textsuperscript{***}}} & 0.86 & 0.17 & -0.04 & 0.49 & 4.18 & 1.26 \\
			\bottomrule
		\end{tabular*}
	}
	\begin{tablenotes}[flushleft]
		\scriptsize
		\item Reported annualized performance measures are the average portfolio return (AV), standard deviation (SD), and Sharpe ratio (SR). We further report the average maximum and minimum weights (MAX, MIN), fraction of negative weights (NEG), gross leverage (LEV), and monthly turnover (TO). All measures are calculated over the 8,841 out-of-sample simple daily returns of the corresponding portfolio. 
		Bold values identify the lowest SD. Asterisks (*, **, ***) indicate statistically significant differences in SD between the approach shown in bold font (best-performing) and each of the remaining approaches; the largest $p$-value across these pairwise tests is used to determine significance at the 10\%, 5\%, and 1\% levels, respectively. $^{\dagger}$ indicates positive values smaller than 0.005.
	\end{tablenotes}
\end{table}
\FloatBarrier

\paragraph{\textit{Different ARMA specifications}}
Table~\ref{tab:perf_gmv_arma_all} illustrates how the portfolio results of the three ARMA-based modeling approaches vary across alternative autoregressive and moving-average lag specifications in the volatility estimation. In line with the motivation presented in Section~\ref{Ch:Pooling}, we consider several ARMA$(p,p)$ specifications. As before, we conduct pairwise significance tests within each approach and find that the choice of $p$ does not have a statistically significant impact on the performance in any of the three approaches. In the univariate case, the annualized SD indicates that the approach based on an ARMA$(1,1)$ outperforms the ARMA$(2,2)$. However, in unreported results, we find that this difference can be attributed primarily to divergent performance during the period surrounding the COVID-19 pandemic. Across all other market phases, performance remains comparable. Results for the univariate ARMA$(3,3)$ are not reported, as estimation proved unstable given the limited sample size.
\begin{table}[htb]
	\centering
	\scriptsize
	\caption{Performance measures for GMV portfolios across different ARMA specifications.}
	\label{tab:perf_gmv_arma_all}
	\resizebox{\textwidth}{!}{%
		\setlength{\tabcolsep}{4pt}
		\begin{tabular*}{\textwidth}{l@{\hspace{53pt}}@{\extracolsep{\fill}}rrrrrrrr}
			\toprule
			\multicolumn{9}{c}{Evaluation period: January 1990 -- February 2025} \\
			\toprule
			Method & AV & SD & SR & MAX & MIN & NEG & LEV & TO \\
			\midrule
			\multicolumn{9}{l}{ARMA(Univ.)-RDCC-NL}
			\\
			\midrule
			ARMA(1,1) & 11.01 & 9.78 & 0.86 & 0.19 & -0.04 & 0.49 & 4.43 & 1.72 \\
			ARMA(2,2) & 10.80 & 10.29 & 0.79 & 0.18 & -0.04 & 0.49 & 4.35 & 1.94 \\
			\midrule
			\multicolumn{9}{l}{ARMA(Ind.)-RDCC-NL} \\
			\midrule
			ARMA(1,1)   & 10.69 & 9.27 & 0.87 & 0.17 & -0.04 & 0.49 & 4.59 & 1.61 \\
			ARMA(2,2)   & 10.71 & 9.29 & 0.87 & 0.16 & -0.04 & 0.49 & 4.59 & 1.69 \\
			ARMA(3,3)   & 10.52 & 9.29 & 0.85 & 0.17 & -0.04 & 0.49 & 4.52 & 1.72 \\
			\midrule
			\multicolumn{9}{l}{ARMA(Mkt.)-RDCC-NL} \\
			\midrule
			ARMA(1,1)   & 10.80 & 9.13 & 0.89 & 0.17 & -0.04 & 0.49 & 4.55 & 1.58 \\
			ARMA(2,2)   & 10.76 & 9.12 & 0.89 & 0.17 & -0.04 & 0.49 & 4.55 & 1.66 \\
			ARMA(3,3)   & 10.72 & 9.08 & 0.89 & 0.17 & -0.04 & 0.49 & 4.50 & 1.67 \\
			\bottomrule
		\end{tabular*}
	}
	\begin{tablenotes}[flushleft]
		\scriptsize
		\item Reported annualized performance measures are the average portfolio return (AV), standard deviation (SD), and Sharpe ratio (SR). We further report the average maximum and minimum weights (MAX, MIN), fraction of negative weights (NEG), gross leverage (LEV), and monthly turnover (TO). All measures are calculated over the 8,841 out-of-sample simple daily returns of the corresponding portfolio. 
	\end{tablenotes}
\end{table}
\FloatBarrier

\subsection{Assessment of volatility forecasts}\label{sec:stat_perf}
For methods that share the same correlation forecast and portfolio constraints, Equation~\eqref{eq:covariance_forecast} 
implies that differences in their estimated covariance matrices and resulting portfolio weights arise
solely from differences in the volatility forecasts. As most of our approaches share the same RDCC-NL correlation forecast,  we focus on two aspects of the volatility forecasts that are particularly important for the performance of our minimum variance portfolios. 
The first is the alignment of the cross-sectional rankings of realized volatilities and their forecasts, which we measure using Spearman's rank correlation coefficient, as detailed in \ref{Ch:Appendix_Corr}. The second is the extent to which forecasts reproduce the relative differences in realized volatility across stocks.

Importantly, a common rescaling of all volatility forecasts does not affect minimum variance portfolio weights. This becomes apparent from the following scale-invariance property of the optimization problem, valid for any constant \(c>0\):
\begin{equation}
	\operatorname*{arg\,min}_{w_t\in\mathcal{W}_t}
	w_t^\top\!\left(c\hat{\Sigma}_t\right)w_t
	=
	\operatorname*{arg\,min}_{w_t\in\mathcal{W}_t}
	w_t^\top\hat{\Sigma}_t w_t,
	\label{eq:covariance_scale_invariance}
\end{equation}
provided that the feasible set \(\mathcal{W}_t\), which is determined by the optimization constraints, is unchanged. 
A common multiplicative error in all volatility forecasts therefore affects the scale of the covariance structure but has no effect on resulting portfolio weights. By contrast, errors that affect the relative relationship between volatilities across stocks can affect the portfolio composition.
Conventional metrics such as the mean squared error combine both sources of error into a single measure and therefore do not distinguish between them. 
We therefore disentangle forecast errors into a ``scale error'' part, reflecting a common cross-sectional deviation of the forecasts from realized volatilities, and a ``dispersion error'' part, reflecting cross-sectional variation of the forecasts around their common deviation. \newline 
We evaluate forecast errors on the log-volatility scale:
\begin{align}
	e_{i,t}^{(m)}
	&=
	\log\!\left(
	\frac{\left(\hat{\sigma}_{i,t}^{(m)}\right)^2}
	{RV_{i,t}^{2}}
	\right),
	&
	\bar e_t^{(m)}
	&=
	\frac{1}{N_t}
	\sum_{i=1}^{N_t}e_{i,t}^{(m)}.
	\label{eq:log_variance_error}
\end{align}
Thus, for example, a forecast variance twice the realized variance yields \(\log(2)\), whereas a forecast variance half the realized variance yields \(-\log(2)\). The log transformation therefore treats reciprocal proportional deviations symmetrically and yields an exact additive decomposition of the mean squared forecast error:
\begin{align}
	\operatorname{\log\text{-}MSE}^{(m)}
	&=
	\frac{1}{T}
	\sum_{t=1}^{T}
	\frac{1}{N_t}
	\sum_{i=1}^{N_t}
	\left(e_{i,t}^{(m)}\right)^2
	\nonumber\\
	&=
	\underbrace{
		\frac{1}{T}
		\sum_{t=1}^{T}
		\left(\bar e_t^{(m)}\right)^2
	}_{\text{scale error}}
	+
	\underbrace{
		\frac{1}{T}
		\sum_{t=1}^{T}
		\frac{1}{N_t}
		\sum_{i=1}^{N_t}
		\left(e_{i,t}^{(m)}-\bar e_t^{(m)}\right)^2
	}_{\text{dispersion error}}.
	\label{eq:log_mse_decomposition}
\end{align}
Notably, the above second-moment identity of the log-MSE mirrors the idea of a bias-variance trade-off, with the scale error capturing a bias-like common deviation and the dispersion error capturing variance-like cross-sectional variation around that deviation. Table~\ref{tab:rank-log-error} reports the two error components together with the overall log-MSE and the average Spearman correlations.
\begin{table}[h!]
	\centering
	\scriptsize
	\caption{Cross-sectional rank accuracy and decomposition of log-variance forecast errors.}
	\label{tab:rank-log-error}
	\resizebox{\textwidth}{!}{%
		\begin{tabular}{lrrrrr}
			\toprule
			\multicolumn{6}{c}{Evaluation period: January 1990 -- February 2025} \\
			\toprule
			Method & Mean $\rho$ & Median $\rho$ & Log-MSE & Scale error & Dispersion error \\
			\midrule
			NL & 0.556 & 0.572 & 1.137 & 0.655 & 0.482 \\
			RDCC-NL & 0.670 & 0.673 & 0.597 & 0.222 & 0.375 \\
			AFM-RDCC-NL & 0.649 & 0.663 & 0.621 & 0.227 & 0.394 \\
			SFM-RDCC-NL & 0.687 & 0.689 & 0.501 & 0.154 & 0.347 \\
			SFM(lrv)-RDCC-NL & 0.683 & 0.684 & 0.505 & 0.155 & 0.351 \\
			SFM(lf)-RDCC-NL & 0.681 & 0.686 & 0.502 & \textbf{0.149} & 0.354 \\
			GDFM-RDCC-NL & 0.694 & 0.695 & \textbf{0.495} & 0.157 & 0.338 \\
			ARMA(Univ.)-RDCC-NL & 0.690 & 0.692 & 0.502 & 0.160 & 0.342 \\
			ARMA(Ind.)-RDCC-NL & \textbf{0.697} & 0.700 & \textbf{0.495} & 0.161 & \textbf{0.335} \\
			ARMA(Mkt.)-RDCC-NL & \textbf{0.697} & \textbf{0.702} & 0.496 & 0.161 & \textbf{0.335} \\
			\bottomrule
		\end{tabular}
	}
	\begin{tablenotes}[flushleft]
		\scriptsize
		\item Mean and median $\rho$ summarize the Spearman rank correlations between forecast and realized volatilities over 421 out-of-sample periods. Log-MSE is decomposed into scale error and dispersion error as defined in Equation~\eqref{eq:log_mse_decomposition}. Higher rank correlations and lower errors indicate better forecast performance. Bold values indicate the best reported values.
	\end{tablenotes}
\end{table}
ARMA(Mkt.) attains the highest median rank correlation, while sharing the highest mean rank correlation with ARMA(Ind.). GDFM and ARMA(Ind.) produce the lowest log-MSE, SFM(lf) the lowest scale error, and the two pooled ARMA specifications the lowest dispersion error. Since common-scale errors do not affect portfolio weights, the combination of accurate cross-sectional rankings and small dispersion errors provides a plausible explanation for the strong portfolio performance of the pooled ARMA models.
 
This pattern is consistent with the model comparison in Section ~\ref{Ch:ModelComparison}. The SFM and GDFM allow greater stock-specific flexibility, which reduces specification bias, in line with their comparatively small scale errors. Pooling imposes stronger commonality and can improve estimation precision, consistent with the lower dispersion errors of the pooled ARMA specifications. Together with their high rank correlations, this suggests that the main advantage of pooling lies in more precise relative volatility forecasts.

\section{Conclusion}
\label{Ch:Conclusion}
We examined the ways in which commonality in realized stock volatilities can be exploited for covariance forecasting and portfolio selection. To this effect, we compared factor models with ARMA models that pool volatility dynamics across stocks and found that pooled ARMA specifications admit factor representations while reducing the parameter burden and restricting the temporal structure of volatility across stocks.
The empirical results favor strong pooling. 
The market-pooled ARMA produces the lowest portfolio volatility in the unconstrained GMV setting. 
For both long-only portfolios and portfolios subject to joint turnover and leverage constraints, the market- and industry-pooled ARMA specifications also remain the best-performing approaches.
Their volatility forecasts exhibit the highest rank correlations and the smallest dispersion errors, whereas the SFM and GDFM achieve lower scale errors.

For minimum variance portfolio selection, both the ranking of stocks by volatility and the relative volatility differences between them matter, while common scaling errors leave portfolio weights unchanged. The results suggest that pooling performs well because it captures both dimensions more accurately.
This pattern is consistent with the bias-variance trade-off highlighted by the model comparison, reflecting a trade-off between lower specification bias from greater factor-model flexibility and lower estimation variance from pooling.

Our results underscore the benefits of pooling dynamic parameters across assets when forecasting volatility with ARMA models. Pooling provides a form of cross-sectional regularization that exploits commonality without requiring the estimation of latent factors. Parameter pooling can, of course, also be applied to other volatility forecasting models; whether, and under what conditions, similar gains can be achieved in those settings remains a question for future research.
\clearpage

\section*{Funding}
This research received no external funding.

\section*{Acknowledgements}
We thank Igor Honig, Felix Kircher, Matteo Barigozzi, Rolf Tschernig, and Daniel Rösch for helpful comments and suggestions. We also thank participants at several university seminars, the 7th Vienna Workshop on High-Dimensional Time Series in Macroeconomics and Finance, and the 2026 EEA-ESEM Congress in Dublin for valuable feedback.

\section*{Data Availability}
Data on stock returns and historical S\&P 500 constituent lists were obtained from LSEG Datastream under a license that does not permit redistribution; they are available to Datastream subscribers.

\section*{Declaration of Competing Interests}
The authors declare that they have no competing interests.

\section*{Declaration on the Use of Generative Artificial Intelligence}
The authors declare that they have used generative artificial intelligence (AI), specifically ChatGPT, Claude and DeepL, to improve the language and ensure the grammatical correctness of this manuscript, as well as to support the development and improvement of code. No original content (including figures or tables) was automatically generated with the help of AI. After using these AI services, the authors reviewed and edited the content of the manuscript as needed and take full responsibility for the content of the publication.

\clearpage
\appendix
\setcounter{figure}{0}%

\section{Figures}\label{app:figures}
\begin{figure}[h!]
	\centering
	\includegraphics[width=\linewidth]{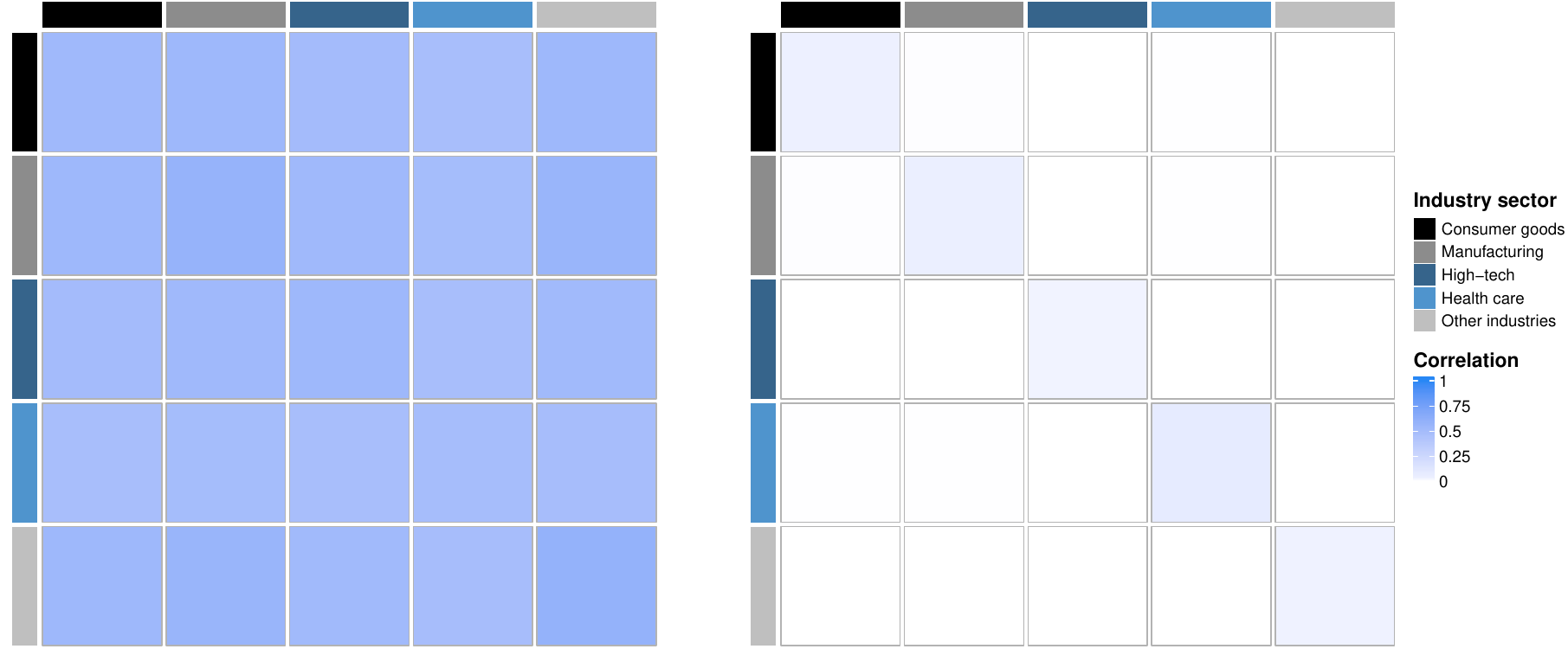}
	\caption{
		 Heatmap of correlations for monthly realized volatilities (\textit{left}) and idiosyncratic volatilities after accounting for cross-sectional variation using three principal components (\textit{right}). Correlations were computed pairwise at the firm level and then averaged by industries.}
		 \alttext{Two heatmaps of average pairwise correlations between firms' monthly
		 	realized volatilities, arranged in a five-by-five grid of industry groups on a
		 	color scale from 0 to 1. Left: all cells show moderate positive correlations,
		 	both within and across industries. Right: after removing three principal
		 	components, cross-industry correlations are close to zero, while some
		 	within-industry correlations along the diagonal remain slightly positive.}
	\label{fig:corr_heatmaps}
\end{figure}
\begin{figure}[t]	
	\centering	
	\includegraphics[width=\linewidth]{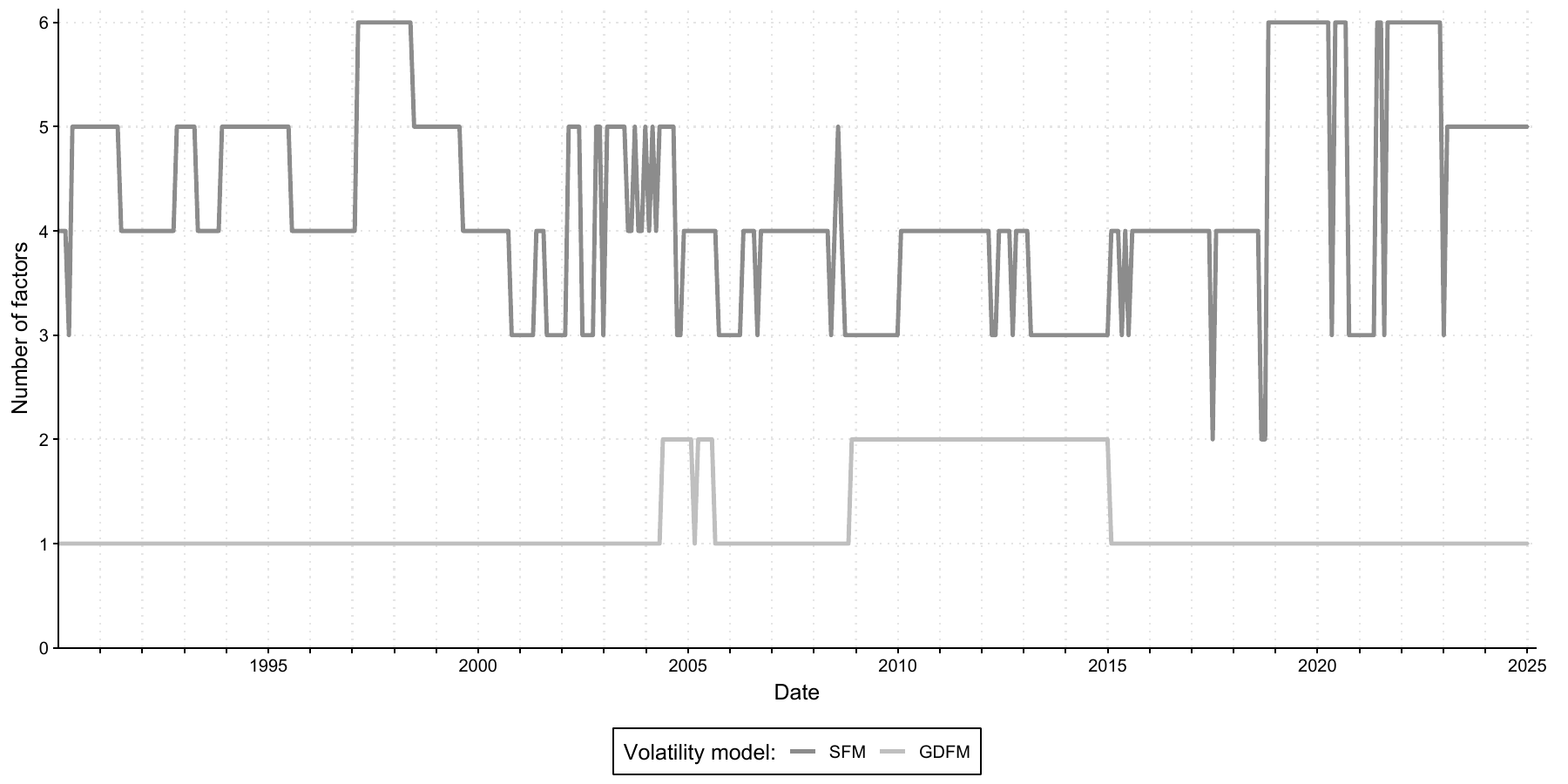}	\caption{ The figure displays the number of factors selected across all rolling estimation windows for the GDFM using the information criterion of \citet{Hallin2007} and for the SFM using the criterion of \citet{Bai2002} together with the penalization technique proposed by \citet{Alessi2010}. The factors determined for the SFM apply analogously to the SFM(lf) and SFM(lrv) specifications.}	
	\label{fig:nof_factors_plot}
	\alttext{Step chart of the number of factors selected in each rolling estimation
		window from 1990 to 2025. The static factor model selects between two and six
		factors, mostly three to five, and changes frequently between windows. The
		dynamic factor model selects one factor in most windows and two factors around
		2004--2005 and from late 2008 to early 2015.}
\end{figure}
\clearpage

\section{Computation of the Dynamic Factor Model}\label{Ch:Appendix_FactorModel}
We implement the one-sided GDFM of \citet{Forni2015,Forni2017}, following the computational procedure for log-volatility panels in \citet{Barigozzi2020}. The method first estimates the spectral density of the volatility panel and reconstructs the spectrum associated with its leading dynamic eigenvalues. Blockwise autoregressive filters then transform the dynamic factor structure into a static representation. Principal components of the filtered panel recover the dynamic factors, while inversion of the autoregressive filters yields the corresponding dynamic loading filters. These estimates are subsequently used to reconstruct and forecast the common component.

The construction depends on the random partition of the cross-section into small blocks. We therefore repeat the estimation for several random permutations and average the resulting factors and loading filters. Because factors and loadings are identified only up to an orthogonal rotation, the estimates obtained from different permutations cannot be averaged directly. After returning the loading matrices to the original asset ordering, we rotate the factors and loadings from each permutation such that the leading $q\times q$ block of the contemporaneous loading matrix is lower triangular with positive diagonal elements. This normalization aligns the factor orientation across permutations without imposing a structural interpretation on the factors.

The implementation requires a spectral bandwidth $m$, the number of dynamic factors $q$, the number of random cross-sectional permutations $\texttt{nreps}$, and a truncation order $p_f$ for the estimated inverse filters. We set
\begin{align}
	m &= \left\lfloor\sqrt{\Tau}\right\rfloor,
	&
	\texttt{nreps} &= 100,
	&
	p_f &= 20.
\end{align}
Because each estimation sample contains $\Tau=120$ monthly observations, the spectral bandwidth is $m=10$. The number of dynamic factors is selected using the $IC^{\Tau}_{2;N}$ criterion of \citet{Hallin2007}, together with its $p_1$ penalty and stability calibration. The block-specific VAR orders are selected by BIC and therefore need not be specified separately. The tuning rules and numerical settings are fixed before the out-of-sample evaluation. The number of dynamic factors and the block-specific VAR orders are determined anew in each estimation window.
We also examined alternative spectral bandwidths, truncation orders, and numbers of random permutations. These alternative specifications produced only minor differences, none of which meaningfully affected our results. The implementation code and detailed sensitivity results are available from the authors upon request. 

\begin{algorithm}[H]
	\caption{Implementation of the generalized dynamic factor model}
	\footnotesize
	\label{alg:gdfm}
	\begin{algorithmic}[1]
				\vspace{0.5em}
		\Statex \textbf{Input:}
		\Statex
		\begin{minipage}{.94\linewidth}
			\begin{itemize}
				\setlength{\itemsep}{0pt}
				\setlength{\parskip}{0pt}
				\setlength{\parsep}{0pt}
				\item Demeaned log-volatility
				$\{v_\tau\}_{\tau=1}^{\Tau}$, with
				$v_\tau\in\mathbb R^{N_t}$.
				\item Spectral bandwidth $m=\lfloor\sqrt{\Tau}\rfloor$.
				\item Bartlett weights $\{\omega_k\}_{k=-m}^{m}$, with $k$ denoting the autocovariance lag.
				\item Maximum number of dynamic factors $q_{\max}=5$.
				\item Number of permutations $\texttt{nreps}=100$.
				\item Inverse-filter truncation order $p_f=20$.
			\end{itemize}
		\end{minipage}
				\vspace{0.5em}
		\Statex \textbf{Output:} 		\vspace{-0.5em}
		\Statex
		\begin{minipage}{.94\linewidth}
			\begin{itemize}
				\setlength{\itemsep}{0pt}
				\setlength{\parskip}{0pt}
				\setlength{\parsep}{0pt}
				\item One-month-ahead forecasts
				$\{\hat v_{i,\Tau+1\mid\Tau}^{\mathrm{GDFM}}\}_{i=1}^{N_t}$.
			\end{itemize}
		\end{minipage}
		\vspace{-1em}
		\Statex \hrulefill
		\Statex \textbf{Step 1: Estimate the spectral density}
		
		\For{$k=0,\ldots,m$}
		\State Compute the sample autocovariance
		$\hat\Gamma_{v,k}
		=(\Tau-k)^{-1}\sum_{\tau=k+1}^{\Tau}
		v_\tau v_{\tau-k}^{\top}$.
		\EndFor
		\State Set
		$\hat\Gamma_{v,-k}=\hat\Gamma_{v,k}^{\top}$
		for $k=1,\ldots,m$.
		\For{$h=-m,\ldots,m$}
		\State Set $\theta_h=2\pi h/(2m+1)$ and evaluate
		$\hat\Sigma_v(\theta_h)
		=\sum_{k=-m}^{m}\omega_k
		\hat\Gamma_{v,k}e^{-ik\theta_h}$.
		\State Obtain the ordered eigenvalues $\{\hat\kappa_i(\theta_h)\}_{i=1}^{N_t}$ and corresponding 
		 normalized eigenvectors $\{\hat p_i(\theta_h)\}_{i=1}^{N_t}$.
		\EndFor
		\State Select $q\leq q_{\max}$ using the
		$IC^{\Tau}_{2;N}$ criterion of \citet{Hallin2007}.
		\stepoutput{%
			The estimated spectral eigenstructure and selected factor dimension $q$.
		}
		\vspace{0.5em}
\algstore{gdfmA}
\end{algorithmic}
\end{algorithm}

\begin{algorithm}[H]
\ContinuedFloat
\caption{Implementation of the generalized dynamic factor model (continued)}
\footnotesize
\begin{algorithmic}[1]
\algrestore{gdfmA}
\Statex \textbf{Step 2: Reconstruct the common spectrum}
		\For{$h=-m,\ldots,m$}
		\State Compute
		$\hat{\Sigma}_{\chi}(\theta_h)
		=
		\sum_{i=1}^{q}
		\hat{\kappa}_i(\theta_h)
		\hat{p}_i(\theta_h)
		\hat{p}_i(\theta_h)^\dagger$,
		where ${}^\dagger$ denotes 
		the conjugate transpose.
		\EndFor
		\For{$k=0,\ldots,m$}
		\State Compute
		$\hat\Gamma_{\chi,k}
		=(2m+1)^{-1}\sum_{h=-m}^{m}
		\hat\Sigma_\chi(\theta_h)
		e^{ik\theta_h}$.
		\EndFor
		\For{$k=1,\ldots,m$}
		\State Set
		$\hat\Gamma_{\chi,-k}=\hat\Gamma_{\chi,k}^{\top}$.
		\EndFor
		\stepoutput{%
			The estimated autocovariances of the common component
			$\{\hat\Gamma_{\chi,k}\}_{k=-m}^{m}$.
		}
							\Statex \hrulefill
		\Statex \textbf{Step 3: Construct blockwise autoregressive filters}
		
		\For{$r=1,\ldots,\texttt{nreps}$}
		\State Randomly permute the $N_t$ volatility series and use the first
		$\mathbf n(q+1)$
		series to form 		\Statex \squad
		$\mathbf n=\lfloor N_t/(q+1)\rfloor$ blocks of size $q+1$.
		\If{$\mathbf n(q+1)<N_t$}
		\State Append the remaining
		$N_t-\mathbf n(q+1)$ series to the final block.
		\EndIf
		
		\For{$b=1,\ldots,\mathbf n$}
		\State Extract from
		$\{\hat\Gamma_{\chi,k}\}_{k=0}^{m}$
		the submatrices associated with
		the series in block $b$.
		\State Select the VAR order
		$p_{b,r}^{\mathrm{VAR}}$ by BIC, estimate the VAR using the
		Yule--Walker equations,
				\Statex \squad \squad and denote its lag polynomial by $A^{(b,r)}(L)$.
		\EndFor
		
		\State Collect the estimated block-specific VAR lag polynomials in
		\Statex \squad
		$A^{(r)}(L)
		=\operatorname{diag}\!\left(
		A^{(1,r)}(L),\ldots,A^{(\mathbf n,r)}(L)
		\right)$,
		with $A^{(r)}(0)=I_{N_t}$.
		
		\State Apply the resulting block-diagonal VAR filter to the
		permuted panel,
		$s_\tau^{(r)}=A^{(r)}(L)v_\tau^{(r)}$.
		
		\vspace{0.5em}
		\stepoutput{%
			The blockwise autoregressive filter $A^{(r)}(L)$ and filtered panel
			$\{s_\tau^{(r)}\}_{\tau=1}^{\Tau}$ for \\ permutation $r$.
		}
		\vspace{0.5em}
				\algstore{gdfmB}
		
		\end{algorithmic}
	\end{algorithm}

\begin{algorithm}[H]
	\ContinuedFloat
	\caption{Implementation of the generalized dynamic factor model (continued)}
	\footnotesize
	\begin{algorithmic}[1]
		\algrestore{gdfmB}
		\Statex \textbf{Step 4: Estimate the unrotated factors}
		
		\State Compute the sample covariance matrix of
		$\{s_\tau^{(r)}\}_{\tau=1}^{\Tau}$.
		\State Let $\tilde H^{(r)}$ contain its $q$ leading normalized
		eigenvectors multiplied
		by $\sqrt{N_t}$.
		\State Estimate the factors as
		$\tilde f_\tau^{(r)}
		=N_t^{-1}\tilde H^{(r)\top}s_\tau^{(r)}$.
		\stepoutput{%
			The unrotated factor estimates
			$\{\tilde f_\tau^{(r)}\}_{\tau=1}^{\Tau}$ for permutation $r$.
		}
				\Statex \hrulefill
		\Statex \textbf{Step 5: Recover the unrotated dynamic loading filters}
		
		\State Compute
		$\tilde\Lambda^{(r)}(L)
		=[A^{(r)}(L)]^{-1}\tilde H^{(r)}$
		and truncate the filter at lag $p_f$.
		\stepoutput{%
			The unrotated dynamic loading filter
			$\tilde\Lambda^{(r)}(L)$ for permutation $r$.
		}
		
		\Statex \hrulefill
		\Statex \textbf{Step 6: Align the factor orientation}
		
		\State Reorder the rows of $\tilde\Lambda^{(r)}(L)$ to match the
		original ordering of the
		volatility series.
		\State Choose an orthogonal matrix $\mathcal R^{(r)}$ such that the
		leading $q\times q$ block
		of $\tilde\Lambda^{(r)}(0)\mathcal R^{(r)}$ 
				\Statex \squad is lower triangular
		with positive diagonal elements.
		\State Set
		$\hat\Lambda^{(r)}(L)
		=\tilde\Lambda^{(r)}(L)\mathcal R^{(r)}$
		and
		$\hat f_\tau^{(r)}
		=\mathcal R^{(r)\top}\tilde f_\tau^{(r)}$.
		\vspace{0.5em}
		\stepoutput{%
			The aligned factor estimates
			$\{\hat f_\tau^{(r)}\}_{\tau=1}^{\Tau}$ and dynamic loading filter
			$\hat\Lambda^{(r)}(L)$ for \\ permutation $r$. 
		}\vspace{1em}
		\EndFor
				\Statex \hrulefill
		\Statex \textbf{Step 7: Aggregate across permutations}
		
		\State Average the estimated factors,
		$\hat f_\tau
		=\texttt{nreps}^{-1}
		\sum_{r=1}^{\texttt{nreps}}\hat f_\tau^{(r)}$.
		
		\For{$\ell=0,\ldots,p_f$}
		\State Average the dynamic loading coefficients,
		$\hat\Lambda_\ell
		=\texttt{nreps}^{-1}
		\sum_{r=1}^{\texttt{nreps}}
		\hat\Lambda_\ell^{(r)}$.
		\EndFor
		
		\State Define
		$\hat\Lambda(L)=\sum_{\ell=0}^{p_f}\hat\Lambda_\ell L^\ell$.
		\State Reconstruct the common component using the truncated filter,
		\Statex 
		$\hat\chi_\tau=\hat\Lambda(L)\hat f_\tau$,
		and set
		$\hat u_\tau=v_\tau-\hat\chi_\tau$.
		\vspace{0.5em}
		\stepoutput{%
			The permutation-averaged factor representation
			$\hat\Lambda(L)$, $\{\hat f_\tau\}$,
			$\{\hat\chi_\tau\}$, and $\{\hat u_\tau\}$.
		}
		\vspace{0.5em}
		\algstore{gdfmC}

\end{algorithmic}
\end{algorithm}

\begin{algorithm}[H]
\ContinuedFloat
\caption{Implementation of the generalized dynamic factor model (continued)}
\footnotesize
\begin{algorithmic}[1]
\algrestore{gdfmC}
		\Statex \textbf{Step 8: Construct the one-month-ahead forecast}
		
		\State Compute the common-component forecast
		$\hat\chi_{\Tau+1\mid\Tau}
		=\sum_{\ell=1}^{p_f}
		\hat\Lambda_\ell\hat f_{\Tau+1-\ell}$.
		\State Forecast each idiosyncratic component
		$\hat u_{i,\Tau+1\mid\Tau}$ using an ARMA$(1,1)$
		model.
		\State Combine both forecasts as
		$\hat v_{i,\Tau+1\mid\Tau}^{\mathrm{GDFM}}
		=\hat\chi_{i,\Tau+1\mid\Tau}
		+\hat u_{i,\Tau+1\mid\Tau}$.
		\vspace{0.5em}
		\stepoutput{%
			The one-month-ahead GDFM forecasts
			$\{\hat v_{i,\Tau+1\mid\Tau}^{\mathrm{GDFM}}\}_{i=1}^{N_t}$.
		}
		\vspace{0.5em}
	\end{algorithmic}
\end{algorithm}

\section{Calculation of Rank Correlation}\label{Ch:Appendix_Corr}
Let \(\hat{\sigma}_{i,t}^{(m)}\) denote the volatility forecast in \(t\) for stock \(i\) under model \(m\), i.e., the \(i\)th element of \(\hat{\sigma}_{\Tau+1\mid\Tau}^{(m)}\) from Equation~\eqref{eq:logexp_fc}. The corresponding cross-sectional ranks are
\begin{align}
	R_{i,t}^{(m)}
	&=
	\left.
	\operatorname{rank}_{j\in\{1,\ldots,N_t\}}
	\!\left(\hat{\sigma}_{j,t}^{(m)}\right)
	\right|_{j=i},
	&
	R_{i,t}^{RV}
	&=
	\left.
	\operatorname{rank}_{j\in\{1,\ldots,N_t\}}
	\!\left(RV_{j,t}\right)
	\right|_{j=i},
\end{align}
where ranks are assigned in ascending order, so that rank one corresponds to the lowest forecast or realized volatility, respectively. Ties are assigned their average rank, and ranks are computed separately for each out-of-sample month. Spearman's rank correlation coefficient is then
\begin{equation}
	\rho_t^{(m)}
	=
	\frac{
		\displaystyle
		\sum_{i=1}^{N_t}
		\left(
		R_{i,t}^{(m)}
		-
		\frac{1}{N_t}\sum_{j=1}^{N_t}R_{j,t}^{(m)}
		\right)
		\left(
		R_{i,t}^{RV}
		-
		\frac{1}{N_t}\sum_{j=1}^{N_t}R_{j,t}^{RV}
		\right)
	}{
		\displaystyle
		\sqrt{
			\sum_{i=1}^{N_t}
			\left(
			R_{i,t}^{(m)}
			-
			\frac{1}{N_t}\sum_{j=1}^{N_t}R_{j,t}^{(m)}
			\right)^2
		}
		\sqrt{
			\sum_{i=1}^{N_t}
			\left(
			R_{i,t}^{RV}
			-
			\frac{1}{N_t}\sum_{j=1}^{N_t}R_{j,t}^{RV}
			\right)^2
		}
	}.
	\label{eq:cross_sectional_rank_correlation}
\end{equation}
Higher values of $\rho_t^{(m)}$ indicate that the forecast more accurately reproduces the cross-sectional ordering of realized volatilities.

\clearpage
\bibliographystyle{chicago}

\bibliography{references.bib}
\clearpage
\theendnotes
\end{document}